\documentclass[twocolumn]{autart}    % Enable this line and disable the 
\usepackage{graphicx,tcolorbox,float,color,subcaption}         
\graphicspath{{Figures/}}
\usepackage{subcaption}  
\usepackage{enumitem}
\usepackage{blindtext}
\usepackage{tikz}
\usetikzlibrary{shapes,arrows}
\usepackage{booktabs,soul,comment,pifont}
\usepackage{amsmath, amssymb, mathtools,sansmath}

\def\calf{{\mathcal{F}}}
\def\calk{{\mathcal{K}}}
\def\calx{{\mathcal{X}}}
\def\cale{{\mathcal{E}}}
\def\rea{\mathbb{R}}

\def\col{\mbox{col}}
\def\nt{N_{\tt traj}}
\def\phil{\phi^{\tt L}}

\usepackage{array}
\newcolumntype{M}[1]{>{\centering\arraybackslash}m{#1}}

\begin{document}

\begin{frontmatter}
\title{Learning Stable Koopman Embeddings for Identification and Control
\thanksref{footnoteinfo}} 

\thanks[footnoteinfo]{A preliminary version was presented in the 2022 American Control Conference \cite{Fan2022Learning}. Corresponding author: Bowen Yi.}

\author[acfr]{Fletcher Fan}\ead{f.fan@acfr.usyd.edu.au},    % Add the 
\author[montreal]{Bowen Yi}\ead{bowen.yi@polymtl.ca},               % e-mail address 
\author[acfr]{David Rye}\ead{david.rye@sydney.edu.au}, % (ead) as shown
\author[acfr]{Guodong Shi}\ead{guodong.shi@sydney.edu.au},
\author[acfr]{Ian R. Manchester}\ead{ian.manchester@sydney.edu.au}

\address[acfr]{Australian Centre for Robotics and School of Aerospace, Mechanical and Mechatronic Engineering, The University of Sydney, NSW 2006, Australia}
\address[montreal]{Department of Electrical Engineering, Polytechnique Montr\'eal \& GERAD, Montr\'eal, Qu\'ebec H3T 1J4, Canada}


\begin{abstract}                          % 
This paper introduces new model parameterizations for learning discrete-time dynamical systems from data via the Koopman operator and studies their properties. Whereas most existing works on Koopman learning do not take into account the stability or stabilizability of the model -- two fundamental pieces of prior knowledge about a given system to be identified -- in this paper, we propose new classes of Koopman models that have built-in guarantees of these properties. These guarantees are achieved through a novel {\em direct parameterization approach} that leads to {\em unconstrained} optimization problems over their parameter sets. {These results rely on the invertibility of the vector fields for autonomous systems and the generalized feedback linearizability (under smooth feedback), respectively.} To explore the representational flexibility of these model sets, we establish the theoretical connections between the stability of discrete-time Koopman embedding and contraction-based forms of nonlinear stability and stabilizability. The proposed approach is illustrated in applications to stable nonlinear system identification and imitation learning via stabilizable models. Simulation results empirically show that the proposed learning approaches outperform prior methods lacking stability guarantees.
\end{abstract}

\end{frontmatter}

%
%======================
\section{Introduction}
%======================
%
Many fundamental phenomena in engineering and science can be described by dynamical systems, making the modeling of dynamical systems a ubiquitous problem across various domains. These models can not only be used to predict future behavior but have also proven effective in planning, estimation, and designing a controller to interact with the real physical world. In general, deriving a model of a dynamical system from first principles may be challenging or even intractable for cases involving complex tasks, such as imitating human behavior. This is where system identification approaches that learn a model from data become useful.

A central consideration for learning algorithms is the model structure. For identifying memoryless input-output mappings, deep neural networks have achieved state-of-the-art results in many domains, such as image classification \cite{KrizhevskyEtAl2012Imagenet}, speech recognition, and cancer detection \cite{sze2017efficient}. In contrast, learning \emph{dynamical} models introduces additional challenges due to the presence of internal memory and feedback. In particular, ensuring the behavioural properties of dynamical models during learning, including stability and stabilizability, is an important aspect that is non-trivial even for linear systems. For example, even if a  physical system is known to be stable, a model learned from data might exhibit instability due to the unavoidable effects of measurement noise, under-modeling, and the challenges of optimization. 

To address this, some recent works aim to impose constraints in terms of prior physical knowledge, specifically using stability constraints as a control-theoretic regularizer for model learning. As summarized in \cite{FICBIL}, there are two main categories for learning dynamical models with stability guarantees: 1) constrained optimization, and 2) learning potential functions via diffeomorphism. Learning algorithms for stable systems have been comprehensively studied for linear systems (see, for example \cite{lacy2003subspace,GillisEtAl2020note,havens2021imitation}); in contrast, the nonlinear counterpart is more challenging, necessitating universal fitting tools to address nonlinearity. Most approaches for nonlinear stable systems employ the Lyapunov method \cite{KhansariBillard2011Learning} and contraction analysis \cite{tobenkin2010convex,UMEMAN}, which yield constrained optimization with scalability issues. Recent research focuses on \emph{directly parameterizing} stable nonlinear systems to formulate unconstrained optimization \cite{Manek2019Learning,revay2023recurrent}. These are more general than the second category outlined in \cite{FICBIL}.

In recent years, there has been a growing interest in the Koopman operator for the analysis, control, and learning of nonlinear systems \cite{mauroy2016global,TakeishiEtAl2017Learning}. As a composition operator, it  characterizes the evolution of observables from a spectral decomposition perspective \cite{Koopman1931Hamiltonian}. Despite being infinite-dimensional, the Koopman operator itself exhibits linearity and proves powerful in addressing various data-driven analysis and prediction problems \cite{han2020deep,TakeishiEtAl2017Learning,KordaMezic2018Linear}. Through the Koopman operator, nonlinear systems can be studied via a spectral decomposition, akin to linear systems analysis. This has huge potential in applying tools from linear systems theory to nonlinear systems, including global stability analysis \cite{mauroy2016global,YiManchester2021equivalence} and a number of linear control methodologies.

In this paper, we focus on Koopman models -- a recently emerging class of models that are both flexible and interpretable -- and Koopman learning frameworks. When learning a Koopman model from data, one attempts to find a finite-dimensional (usually approximate) representation of the Koopman operator, which amounts to a linear matrix along with a mapping that transforms the original state space of the system to a so-called Koopman-invariant subspace. As mentioned above, it is important to consider model stability and stabilizability during learning.  {However, there are relatively few works, and it has not been fully addressed in existing Koopman learning research. \cite{MamakoukasEtAl2020Learning} employs constrained optimization to search for Hurwitz system matrices to approximate Koopman operator, but this approach is not computationally scalable for high-dimensional systems. The recent work \cite{PanDuraisamy2020Physics} introduces a Koopman learning framework for continuous-time systems with stability guarantees and uncertainty quantification. It relies on \emph{diagonalizable} matrices for the nonlinear reconstruction of observables from Koopman eigenfunctions, which, while structured, imposes a relatively conservative constraint. Additionally, \cite{MITetal} utilizes quadratic programming to enforce Lyapunov stability; however, it does not ensure asymptotic stability.}

The paper aims to address the aforementioned challenges with the main contributions below:
\begin{enumerate}
    \item[1.] We provide a novel parameterization to the stable Koopman model set, which is \emph{unconstrained} in its parameters, allowing for efficient and ``plug-and-play'' optimization by leveraging software tools for automatic differentiation (autodiff).

    \vspace{.3em}
    
    \item[2.]  For nonlinear discrete-time systems, we prove the equivalence between the Koopman and contraction criteria for stability analysis, extending our earlier work  \cite{YiManchester2021equivalence} to the discrete-time context. Such an equivalence is practically useful in proposing a novel Koopman learning framework that is capable of learning  most stable autonomous systems under some technical assumptions.

    \vspace{.3em}
    
    \item[3.] The proposed Koopman model set is extended to the generalized feedback linearizable systems, for which we  develop unconstrained optimization but simultaneously impose the stabilizability constraint to the model set. These results are applied to imitation learning (i.e. learning a control policy from demonstrations) incorporating regularization to guarantee closed-loop stability.

\end{enumerate}
Compared to the preliminary conference version \cite{Fan2022Learning}, this paper provides the full proof of Theorem \ref{thm:1}. Additionally, we extend the main idea to nonlinear control, forming the basis for the stabilizable Koopman model and the imitation learning framework in Section \ref{sec:5}.  

{\em Notation.} All mappings and functions are assumed sufficiently smooth. %Given $f: \rea^n \to \rea^m$, we denote the gradient operator $\nabla f:= ({\partial f \over \partial x})^\top$. 
$\lambda_{\tt min}(\cdot)$ and $\lambda_{\tt max}(\cdot)$ respectively represent the smallest and largest eigenvalues of a symmetric matrix. Given a matrix $A\in \rea^{n\times m}~(n>m)$, $A^\bot \in {\rea^{(n-m) \times n}}$ represents a full-rank left annihilator such that $A^\bot A=0$.
 We use $|\cdot|$ to denote the standard Euclidean norm, i.e. $|x| = \sqrt{x^\top x}$. When clear from context, we may simply write $x(t)$ as $x_t$, omitting the arguments of mappings and functions, and use $\tilde{x}_t$ to represent the measured data corresponding to the true state $x_t$ at time $t$. We use $\circ$ to represent function composition.

 %========================
\section{Preliminaries}
 %========================

This section presents some preliminaries on the Koopman operator and contraction analysis. Consider the discrete-time autonomous system in the form
\begin{equation}\label{eqn:sys_auton}
x_{t+1} = f(x_t)
\end{equation}
with the state $x \in \mathbb{R}^n$, and a smooth vector field $f:\mathbb{R}^n \to \mathbb{R}^n$. Note that the smoothness of vector fields is a standard assumption in nonlinear control theory \cite{ISIbook,khalil2002nonlinear}, which is typically satisfied by most physical models encountered in engineering practice.

The Koopman operator, originally proposed in \cite{Koopman1931Hamiltonian}, has discrete-time version defined as follows.%, provides a simple and effective way to analyze nonlinear systems. 
\begin{defn}\label{def:koopman}\rm
Let $\calf $ be the observable space of scalar functions $\rea^n \to \mathbb{C}$. For the system \eqref{eqn:sys_auton}, the Koopman operator $\mathcal{K}: \mathcal{F} \to \mathcal{F}$ is defined by $\mathcal{K}[\varphi] := \varphi \circ f$
    % \begin{equation}
    % \label{koopman}
    % \mathcal{K}[\varphi] := \varphi \circ f
    % \end{equation}
    for $\varphi \in \mathcal{F}$. 
\end{defn}

Since the Koopman operator acts on a functional space, it is inherently infinite-dimensional. However, its linearity enables powerful methods for analyzing, controlling, and learning nonlinear systems. In many cases, a computationally tractable representation can be obtained by finding a finite set of observables that span an invariant subspace. This allows the operator's action to be represented as a matrix restricted to the subspace using the chosen bases.

\begin{defn}\rm\label{def:koopman_invariance}
    A Koopman-invariant subspace is defined as $\mathcal{G} \subset \calf$ such that $\calk [\varphi] \in \mathcal{G},~ \forall  \varphi \in \mathcal{G}$. 
\end{defn}

If a Koopman-invariant subspace $\mathcal{G}$ is spanned by a finite set of observables $\lbrace\varphi_k\rbrace_{k=1}^N$ with $N \in \mathbb{N}_+$, any function $f \in \mathcal{G}$ can be represented as $f(x) = \sum_{j=1}^N k_j \varphi_j$ with some scalars $k_j$. In some works, this is referred to as the mapping $\phi = \col(\varphi_1, \ldots, \varphi_N)$ as a Koopman embedding of the system \eqref{eqn:sys_auton}\footnote{This is a restrictive assumption for nonlinear systems. In the literature, Koopman embedding is often approximated on a non-invariant set and learned using deep models \cite{TakeishiEtAl2017Learning}.}. Furthermore, if $\phi$ is a surjection, then the original nonlinear system \eqref{eqn:sys_auton} is topologically semiconjugate to a linear system via the coordinate transformation $x \mapsto z=\phi(x)$. More specifically, if $\phi$ is a homeomorphism, it becomes topologically conjugate \cite[Ch. 10]{DEV}. In the topological sense, an injective continuous map $\phi$ is an embedding if $\phi$ yields a homeomorphism between $\rea^n$ and $\phi(\rea^n) \subset \rea^N$. In this paper, we employ the term of ``Koopman embedding'' in the topological sense, which can be viewed as implicitly assuming Koopman invariance; see Section \ref{sec:Koopman_invariance}.

Contraction, also known as incremental exponential stability (IES), is a strong form of stability: if a given system is contracting, any two trajectories ultimately converge to each other \cite{LohmillerSlotine1998Contraction}. Contraction analysis provides another ``exact and global linearization'' way to study nonlinear stability by analyzing the stability of the linear time-varying (LTV) differential system
\begin{equation}
\label{LTV:1}
\delta x_{t+1} = {\partial f \over \partial x}(x_t) \delta x_{t}
\end{equation}
along all feasible trajectories. The new variable $\delta x$, living on the tangent space of the original system \eqref{eqn:sys_auton}, represents the infinitesimal displacement among trajectories. It has shown success in a series of constructive problems for nonlinear systems, including controller synthesis \cite{lohmiller2000control,manchester2017control}, observer design \cite{SANPRA,yi2021reduced}, and learning algorithms \cite{DAVetal,revay2023recurrent,singh2021learning}. We briefly recall the discrete-time definition of contraction as follows.

\begin{defn}
\label{def:contraction}\rm
Given the nonlinear system \eqref{eqn:sys_auton}, if there exists a uniformly bounded metric $M(x)$, {i.e.} $a_1 I_n \preceq M(x) \preceq a_2 I_n$ for some $a_2\ge a_1 >0$, guaranteeing
\begin{equation}
\label{cond:contraction}
{\partial f\over \partial x}(x_t)^\top M(x_{t+1}) {\partial f\over \partial x}(x_t) - M(x_t) \preceq - \beta M(x_t ), 
\end{equation}
with $\beta \in (0,1)$, then the given system is contracting.
\end{defn}

Intuitively, consider a quadratic Lyapunov-like function $V(x,\delta x)=\delta x^\top M(x) \delta x$ on \emph{tangent bundles}. From \eqref{LTV:1}-\eqref{cond:contraction}, it leads to $V(x_{t+1}, \delta x_{t+1}) \le (1-\beta) V(x_t, \delta x_t)$ with $1-\beta \in (0,1)$, and thus $V$ decreases over time, uniformly along all feasible trajectories of $x_t$. A central result of contraction analysis is that, for contracting systems, all trajectories converge exponentially to a single trajectory. That is, for any trajectories $x^a,x^b$ and for some $a_0>0$
\begin{equation}
    |x^a_t - x^b_t| \le a_0\beta^t |x^a_0 - x^b_0|.
\end{equation}
Given the similarity between the contraction and Koopman approaches, our paper \cite{YiManchester2021equivalence} establishes their equivalence for nonlinear stability analysis in the context of \emph{continuous-time} systems.

%
%=======================
\section{Motivations and Problem Set}
\label{sec:3}
%=======================
%

In this paper, we are concerned with the discrete-time nonlinear autonomous system \eqref{eqn:sys_auton} and the control system
\begin{equation}
\label{eq:syst}
x_{t+1} = f(x_t) + g(x_t) u_t,
\end{equation}
but the dynamics is assumed \emph{unknown}, with the state $x\in \rea^n$, the input $u\in \rea^m$, and the vector fields $f:\rea^n \to \rea^n$ and $g: \rea^n \to \rea^{n \times m}$. When there is no external input, i.e. $u \equiv 0$, the control model \eqref{eq:syst} degrades into the autonomous system as introduced in \eqref{eqn:sys_auton}.

Suppose $N_{\tt traj}$ data samples $\mathcal{E}_{\tt D}:=\{\tilde x_t, \tilde u_t\}_{t=1}^{\nt}$ are used for model identification and learning a stabilizing controller, in which $\tilde x, \tilde u$ represents the measured noisy data of $x,u$ generated by the system \eqref{eq:syst} over time. The fundamental question in system identification is to use the dataset $\cale_{\tt D}$ to approximate the vector fields $f,g$, denoted as $\hat f, \hat g$, in some optimal sense. Sometimes, it is necessary to impose additional \emph{constraints} based on prior physical knowledge, such as stability and stabilizability \cite{mania2022active}. We may compactly write as $(\hat f, \hat g) \in \cale_{\tt M}$ with the set $\cale_{\tt M}$ characterizing these constraints. The system identification problem is generally based on minimizing a cost function
\begin{equation}
   \min_{(\hat f, \hat g) \in \cale_{\tt M}} \ J(\cale_{\tt D}, \hat f , \hat g).
\end{equation}
Given a data set, the main considerations of nonlinear system identification are the parameterization of nonlinear functions $\hat f, \hat g$, the selection of the cost function $J$, and specific optimization algorithms. 

This paper proposes two model parameterizations: the stable Koopman model and the stabilizable Koopman model. The main theoretical problem we are interested in is how to parameterize these model sets that are unconstrained in parameters. This endeavor is motivated by and finds practical applications in the following.

{\bf Motivating Applications:}
Given the dataset $\mathcal{E}_{\tt D}$ and a cost function $J(\cale_{\tt D}, \hat f, \hat g)$, solve the following problems.
\begin{itemize}
    \setlength\itemindent{0pt} % Adjust the left margin
    \item[P1:] (Learning stable autonomous systems)
    Considering a contracting system for the case $u\equiv 0$, learn an approximate model $\hat f$ from the dataset $\cale_{\tt D}$ that is generated by the system, by minimizing simulation errors,    
    while ensuring that the identified dynamics $x_{t+1} = \hat f(x_t)$ is contracting.

\vspace{.4em}
    
    \item[P2:] (Imitation learning) Considering the dataset $\cale_{\tt D}$ generated from an asymptotically stabilizable system \eqref{eq:syst} via smooth static feedback, learn a static feedback $u = \rho(x)$  approximating the data and concurrently guaranteeing that the closed loop $x_{t+1} =  f(x_t) + g(x_t) \rho(x_t) $ is contracting. 
\end{itemize}
We will address the above motivating applications in Sections \ref{sec:4} and \ref{sec:5}, respectively. Note that determining the functions $f$ and $g$ requires solving an infinite-dimensional optimization problem. To make this tractable, we parameterize the functions, i.e.
$
\hat f(x,\theta), \ \hat g(x,\theta)
$
using some basis functions that may be selected as polynomials, neural networks or many others. The theoretical question therein is how to introduce parameterizations to guarantee stability and stabilizability properties for the proposed model sets.

%
%=======================
\section{Learning Stable Koopman Embeddings}
\label{sec:4}
%=======================
%

In this section, we focus on the autonomous case, introducing a stable model class that covers all contracting systems under some technical assumptions and studying its equivalent parameterization. Based on them, we propose an algorithm to learn stable Koopman embeddings. 

\subsection{Stable Koopman model class}

\begin{center}
    \fbox{ \parbox { 0.88\linewidth} 
{
\small
{\bf M1.} Stable Koopman Model ($A, \phi, \phil$):
\begin{equation} \label{eqn:model}
    \begin{aligned}
    z_0& ~=~ \phi(x_0) \\
    z_{t+1} &~=~ A z_t, \;t = 0, 1, ... T, \\
    x_t &~=~ \phil(z_t) \quad \forall t,
    \end{aligned}
\end{equation}

in which 1) $z\in \rea^N$ ($N \ge n$) is a lifted internal variable; 2) $A$ is Schur stable; and 3) $\phi$ has a left inverse satisfying $\phil(\phi(x)) = x$, $\forall x$.
}
}
\end{center}
\vspace{-0.1cm}

Let us consider a Koopman model class for discrete-time autonomous systems in the form of \eqref{eqn:sys_auton}. We define a Koopman model for this system in the above table. Here, the left invertibility of $\phi$ implies $ x_t= \phil(z_t)$.

\subsubsection{Stability criterion for Koopman models}

The following theorem is a discrete-time version of the main results in \cite{YiManchester2021equivalence}, showing the equivalence between the Koopman and contraction approaches. As a consequence, it illustrates the Koopman model class {\bf M1} covers all the contracting discrete-time autonomous systems under some technical assumptions on the vector field $f$.
\begin{assum}\rm\label{assm:1}
 The vector field $f$ is invertible and its inverse $f^{-1}$ is continuous satisfying $|x| \le c_1 + c_2|f(x)|$ for some $c_1,c_2\in \rea_+$ and $\|{\partial f \over \partial x}(x)\|< \min\{1, c_2^{-1}\}$. 
\end{assum}

% Thus, the model class is capable of providing sufficient degrees of freedom for learning discrete-time nonlinear systems. 
We have the following.

\begin{thm}\rm
	\label{thm:1}
	Consider the system \eqref{eqn:sys_auton}. Assume that there exists a $C^1$-continuous mapping $\phi: \rea^n \to \rea^N$ with $N \ge n$ such that
	\begin{itemize}
		\item[D1:] There exists a Schur stable matrix $A \in \rea^{N\times N}$ satisfying the algebraic equation
		\begin{equation}
		\label{AEq}
		{\phi\circ f}  - A \phi = 0, \quad \forall x\in \rea^n.
		\end{equation}

  %       \begin{equation}
		% \label{AEq}
		% \red{\calk(\phi) = A \phi, \quad \forall x\in \rea^n.}
		% \end{equation}
        
		% \begin{equation}
		% \label{AEq}
		% \red{\calk(\phi)\circ f  - A \calk(\phi) = 0, \quad \forall x\in \rea^n.}
		% \end{equation}
        
		%%$\Phi:= (\nabla \phi)^\top$
		\item[D2:] ${\partial \phi \over \partial x}$ has full column rank, and $({\partial \phi \over \partial x})^\top {\partial \phi \over \partial x}$ is uniformly bounded.
	\end{itemize}
	Then, the system is contracting with the contraction metric $ M(x):= {\partial \phi \over \partial x}(x)^\top P {\partial \phi \over \partial x}(x)$, where $P$ is any positive-definite matrix satisfying $P - A^\top P A \succ 0$. 
    
    Conversely, if the system \eqref{eqn:sys_auton} is contracting and satisfies Assumption \ref{assm:1}, then in any forward invariant\footnote{For a complete system \eqref{eqn:sys_auton} with the solution $X(x,k)$, a set $\calx$ is said to be forward invariant if whenever $x_0\in\calx$ and $j\in \mathbb{Z}_+$, we have $X(x_0,j) \in \calx$ \cite[pp. 198]{SASbook}.} compact set $\calx \subset \rea^n$ for the dynamics \eqref{eqn:sys_auton}, there exists a $C^1$-continuous mapping $\phi: \rea^n \to \rea^N$ verifying {D1}-{D2}.
\end{thm}

\begin{pf}
	($\Rightarrow$) We need to verify the contraction condition \eqref{def:contraction} from the Koopman conditions D1 and D2.

From D1 (Schur stability of $A$), there exists a matrix $P = P^\top \succ 0$ satisfying the Lyapunov condition
	\begin{equation}
	\label{lya_eq}
	P - A^\top P A \succ I \succeq \rho I ,
	\end{equation}
for some $\rho\in (0,1]$. Invoking the $C^1$-continuity of $\phi$ and $f$, we calculate the partial derivative of \eqref{AEq}, obtaining
 \begin{equation}
 \label{eq:AEq2}
 {\partial \phi \over \partial x}(f(x)) {\partial f \over \partial x}(x) = A {\partial \phi \over \partial x}(x).
 \end{equation}
 %For convenience, we define $F:= {\partial f \over \partial x}$. 
From $x_{t+1} = f(x_k)$, the above can be rewritten as
	\begin{equation}
    \label{eq:AEq3}
	    	{\partial \phi \over \partial x}(x_{t+1}) {\partial f \over \partial x}(x_t) = A {\partial \phi \over \partial x}(x_t).
	\end{equation}
	Due to the full rank of ${\partial \phi \over \partial x}$ and \eqref{lya_eq}, it follows that 
	\begin{equation}
	\label{phiQphi}
	{\partial \phi \over \partial x}(x)^\top(P - A^\top P A){\partial \phi \over \partial x}(x) \succ {\partial \phi \over \partial x}(x)^\top \rho I {\partial \phi \over \partial x}(x).
	\end{equation}
	Then, by substituting \eqref{eq:AEq3}, 
	we have 
	\begin{equation} \label{eqn:contraction_cond_1}
	\begin{aligned}
	& ~{\partial \phi \over \partial x}(x_t)^\top P {\partial \phi \over \partial x}(x_t) -  {\partial f \over \partial x}(x_t)^\top {\partial \phi \over \partial x}(x_{t+1})^\top P {\partial \phi \over \partial x}(x_{t+1})  {\partial f \over \partial x}(x_t)
	\\
	\succ &~ \rho {\partial \phi \over \partial x} (x_t)^\top  {\partial \phi \over \partial x} (x_t)
    \\
     \succeq & ~\beta  {\partial \phi \over \partial x}(x_t)^\top P {\partial \phi \over \partial x}(x_t),
	\end{aligned}
	\end{equation}
    with $\beta:={\rho \over \lambda_{\tt max}(P)}$. We choose $M(x) := {\partial \phi \over \partial x}(x)^\top P {\partial \phi \over \partial x}(x)$. From D2 and $P\succ 0$, the selected metric $M(x)$ satisfies the uniform boundedness requirement in Definition \ref{def:contraction}. Substituting into \eqref{eqn:contraction_cond_1} leads to
	\begin{equation}
	M(x_t) -  {\partial f \over \partial x}(x_t)^\top M(x_{t+1}){\partial f \over \partial x}(x_t) \succ \beta M(x_t).
	\end{equation}
	By selecting $\rho \in (0,1]$ sufficiently small, we can guarantee that $\beta \in (0,1)$. This is exactly the contraction condition for the system \eqref{eqn:sys_auton} with the contraction metric $M$.
	
	($\Leftarrow$) This part is to show that a contracting system satisfies the Koopman conditions D1 and D2 in any invariant compact set $\calx \subset \rea^n$. The key step for D1 is to construct a feasible solution to the algebraic equation \eqref{AEq}, which is motivated by the technical results in \cite{BRIetal}.
 
 For the given system, we directly apply the Banach fixed-point theorem, concluding the existence of a unique fixed-point $x_\star \in \calx$, i.e. $f(x_\star) = x_\star$.\footnote{The IES condition \eqref{cond:contraction} implies that for any $x_a,x_b \in \calx$ we have
 $
    d_M(f(x_a),f(x_b))^2 
      \le \beta d_M(x_a, x_b)^2,
 $
 with $\beta\in(0,1)$ and $d_M(\cdot, \cdot)$ the distance associated with the metric $M(x)$.
 Therefore, the mapping $f$ is a contraction mapping in a Banach space, and we may then apply \cite[Appendix B]{khalil2002nonlinear}.
 } To construct an embedding $\phi$, we parameterize as
	$
	\phi(x) = x + T(x),
	$
 for the particular case $N=n$, 
	with a new mapping $T: \rea^n \to \rea^n$ to search. Then, the \eqref{AEq} becomes
	\begin{equation}
 \label{eq:al_T1}
	    T(f(x))+ f(x) = A x + A T(x).
	\end{equation}
	Let us fix $A
 = {\partial f \over \partial x} (x_\star)$. From the contraction assumption, we have 
	$
	M_\star - A^\top M_\star A \succeq \beta M(x^\star)
	$
with $M_\star := M(x_\star) \succ 0$, and thus $A$ is Schur stable. By defining 	
    $
	H(x):= A x - f(x),
	$
 the algebraic equation \eqref{eq:al_T1} becomes 
	\begin{equation}
	\label{AE:T}
	T(f(x)) = A T(x) + H(x).
	\end{equation}
Note that \eqref{AE:T} exactly coincides with the algebraic equation in the Kazantzis-Kravaris-Luenberger (KKL) observer for nonlinear discrete-time systems in
	\cite[Eq. (7)]{BRIetal}. In our case, the function $H$ is continuous and, following \cite[Thm. 2]{BRIetal}, we have a feasible solution to \eqref{AE:T}:
	\begin{equation}
	\label{solution}
	T(x) = \sum_{j=0}^{+\infty} A^j  H(X(x,-j+1)),
	\end{equation}
which is well-defined due to $\| {\partial f \over \partial x}(x_\star)\| <\min\{1, c_2^{-1}\}$, with the definition for $j \in \mathbb{N}_+$
	$$
	X(x,j) = \underbrace{f\circ f\circ \cdots \circ f}_{j~\mbox{\small times}}(x)
	, \quad
	X(x,-j) = (f^{\dagger})^j(x).
	$$
Note that the calculation of $X(x,-j)$ requires backward completeness and invariance in $\calx$. Since contracting systems generally cannot guarantee such invariance, we modify the backwards map $f^{-1}$ as $f^\dagger(x) = \mu(x)f^{-1}(x) + [1-\mu(x)]x$ with
 $$
	\mu(x) = \left\{
	\begin{aligned}
	1, \quad &\mbox{if~~} x\in \mathcal{X} + k_1
 \\
 \mu_f(x), \quad & \mbox{if~~} x\in \mathcal{X} + k_2 \backslash \mathcal{X} + k_1
	\\
	0, \quad & \mbox{if ~~} x\notin \mathcal{X}+k_2
	\end{aligned}
	\right.
	$$
for some $k_2> k_1 >0$, where $\mathcal{X}+k_1$ represents the set of points that lie within in the distance $k_1>0$, and $\mu_f(x)$ is any locally Lipschitz function such that $\mu$ is $C^1$-continuous \cite{TRABER}.

	Now, we consider a candidate Koopman embedding $\phi^0(x) := x+ T(x)$ with $T$ defined above satisfying D1 $\forall x\in \calx$. However, the condition D2 does not necessarily hold, and we need to modify $\phi^0$. By considering the evolution of the trajectories in the $x$-coordinate and a lifted coordinate defined as $z:=\phi(x)$, respectively, we have
	$$
	z(t_x) = \phi^0(x(t_x))= \phi^0(X(x,t_x)) = A^{t_x} \phi^0(x),
	$$
	with $t_x\in \mathbb{N}_+$, thus satisfying
	$
	\phi^0(x) = A^{-t_x}\phi^0(X(x,t_x)).
	$
	Then, we modify the candidate embedding $\phi^0$ into
	\begin{equation}
	\label{phi}
	\phi(x) := A^{-t_x}[X(x,t_x) + T(X(x,t_x))]
	\end{equation}
	with a sufficiently large $t_x \in \mathbb{N}_+$.
	
	Finally, let us check conditions {D1} and {D2}. For D1,
	$$
	\begin{aligned}
	\phi\circ f(x) & = A^{-t_x} \phi^0 \circ X(f(x),t_x)
	\\
	& =  A^{-t_x} \phi^0 \circ f \circ X(x, t_x)
	\\
	& 
	= A^{-t_x} \cdot A\phi^0 \circ X(x,t_x)
	%\\
	%& 
	%= A\cdot A^{-t_x} \phi^0(X(x,t_x))
	\\
	& 
	= A\phi(x),
	\end{aligned}
	$$ 
	where in the second equation we have used the fact 
	$$
	X(f(x),t_x) =  \underbrace{f\circ f\circ \cdots \circ f}_{(t_x+1)~\mbox{\small times}} 
	=
	f(X(x,t_x)).
	$$ 
	Therefore, $\phi$ defined in \eqref{phi} satisfies D1. Regarding {D2}, the Jacobian of $\phi$ is given by
	$$
	{\partial \phi \over \partial x}(x) = A^{-t_x} \left[
	I + {\partial T \over \partial x}(X(x,t_x) )
	\right]
	{\partial X \over \partial x}(x,t_x).
	$$
	Since ${\partial X \over \partial x}$ is full rank and
	$
	H (x^\star) = 0$, ${\partial H \over \partial x}(x_\star) =0,
	$
	we have ${\partial T \over \partial x}(x_\star)=0$. If $t_x \in \mathbb{N}_+$ is sufficiently large, the largest singular value of ${\partial T \over \partial x}(X(x,t_x))$ is tiny, and the identity part of $\phi$ dominates ${\partial \phi \over \partial x}$. Hence, $\phi$ satisfies the condition D2 that ${\partial \phi \over \partial x}$ has full column rank, and the uniform boundedness can be obtained in a compact set. %We complete the proof.
 \hfill $\square$
\end{pf}

\begin{figure}[!htp]
\begin{center}\scriptsize
% Define new styles for the blocks
\tikzstyle{block} = [draw, fill=blue!20, rectangle, 
    minimum height=2.5em, minimum width=6em, rounded corners]
\tikzstyle{input} = [coordinate]
\tikzstyle{output} = [coordinate]
\tikzstyle{arrow} = [thick,->,>=stealth]
\tikzstyle{title} = [font=\bfseries]

\begin{tikzpicture}[auto, node distance=3cm,>=latex']
    % We start by placing the blocks
    \node [input, name=input] {};
    \node [block, , fill=orange!15, right of=input, node distance=2cm,text width=6.5em, minimum width=3em, text centered] (SO) {\scriptsize {\sansmath $\rm Koopman$} {\sansmath $\rm Embedding$} \\$\phi$};
    \node [block, fill=green!15, right of = SO, node distance=3cm, text width=6.5em, minimum width=3em, text centered] (DE) {\scriptsize {\sansmath $\rm Linear~System$}  {\sansmath $\rm Identification$}    $z_{t+1} = Az_t$};
    \node [right of = DE] (output){};
    \node [block, , fill=orange!15, right of = DE , node distance=3cm, text width=5em, minimum width=3em, text centered] (PI) {\scriptsize {\sansmath $\rm Left $}\\ {\sansmath $\rm Inverse$} \\ $\phil$};
    \draw [->] (SO) -- node [name = y1] {} (DE);
    \draw [->] (DE) -- node [name = y1] {} (PI);
\end{tikzpicture}
\end{center}
    \caption{The proposed model class {\bf M1}: Use linear system identification approaches to learn nonlinear models.}
    \label{fig:scheme}
\end{figure}
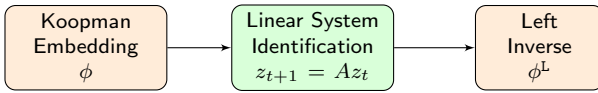

\begin{rem}\rm 
    The above shows the equivalence between Koopman and contraction approaches (i.e. the theoretical conditions) for stability analysis of discrete-time systems. This resembles the results for continuous-time systems in \cite{YiManchester2021equivalence}. %In contrast, we limit ourselves to discrete-time time-invariant systems in this paper.
    Intuitively, this equivalence implies generality and flexibility of the proposed model class {\bf M1}, i.e., it covers all contracting nonlinear systems under some technical assumptions. 
   If the model \eqref{eqn:sys_auton} is derived by discretizing a continuous-time system, the invertibility of $f$ in Assumption \ref{assm:1} can be guaranteed.
    While the construction of the Koopman mapping $\phi$ is an existence result, this shows the potential to use linear system identification techniques to learn a nonlinear model; see Fig. \ref{fig:scheme} for its basic idea. We will pursue it in the next subsection.
\end{rem}

\begin{rem}\rm
For autonomous (i.e. time-invariant unforced) systems, contraction is equivalent to exponential stability within any \emph{compact invariant subset} of the domain of attraction; however, this may not hold over the entire state space. The paper \cite{mauroy2016global} proposes a similar equivalence between global asymptotic stability and the Koopman stability criteria for continuous-time systems. It is the authors’ opinion that contraction provides a more natural and fundamental form of nonlinear stability to link to Koopman representations than asymptotic stability; see \cite{YiManchester2021equivalence} for a comprehensive discussion. 
\end{rem}

\begin{rem}\rm
The above theorem shows that in theory, lifting with excessive coordinates $(N> n)$ is unnecessary to obtain a linear system for a particular class of nonlinear systems, i.e. contracting systems. Similar results are also obtained in \cite{LANMEZ} for Schur stable systems. However, overparameterizing with $N\ge n$ may still be useful for black-box learning as we show empirically via simulations. The condition D2 makes $\phi$ locally injective, thus being an embedding that allows us to lift the given nonlinear system to a linear stable dynamics. %Therefore, we use the terminology ``Koopman embedding''.
\end{rem}

%============

\subsection{Parameterization of stable Koopman models}
\label{sec:412}

In the Koopman model {\bf M1}, we need to identify three components: the stable matrix $A$, the mapping $\phi$, and its left inverse $\phil$. It is necessary to parameterize them to make the approach computationally tractable. 
A key feature of the matrix $A$ is Schur stability, for which there are several equivalent conditions, e.g. the Lyapunov inequality \eqref{lya_eq} and the parameterization in \cite{GillisEtAl2020note}. However, these constraints are non-convex thus yielding heavy computational burden. To address this, we introduce an \textit{unconstrained} parameterization of stable $A$, a special case of the direct parameterization in \cite{revay2023recurrent}. 

\begin{prop} \label{prop:stable_param}\rm 
	Consider a matrix $A\in \rea^{N\times N}$ parameterized as $A(L, R)$\footnote{To simplify the presentation, we use $A$ to represent both a matrix and the parameterization function.}
	\begin{equation} 
    \label{eqn:A_param}
	A(L, R) = 2 (M_{11} + M_{22} + R - R^\top)^{-1}M_{21},
	\end{equation}
	where $M_{ij}$ ($i,j =1,2$) are blocks in 
	\begin{equation} \label{eqn:M}
	M := \begin{bmatrix} M_{11} & M_{12} \\ M_{21} & M_{22} \end{bmatrix} = LL^\top + \epsilon I,
	\end{equation}
	with $\epsilon$ a positive scalar, $L \in \rea^{2N \times 2N}$, and $R \in \rea^{N\times N}$. Then, the matrix $A(L,R)$ is Schur stable. Conversely, for any Schur stable matrix $A$, we can always find $L,R$ and $\epsilon$ to parameterize it in the form of \eqref{eqn:A_param}.
\end{prop}
\begin{pf} 
	($\Rightarrow$)
	Let $E = {1\over 2}(M_{11} + M_{22} + R - R^\top)$, $F = M_{21}$ and $P = M_{22}$. Then, we have $A(L, R) = E^{-1}F$ and 
	\begin{equation} \label{eqn:implicit_mat}
	M = \left[\begin{array}{c|c}
	E + E^\top - P & F^\top \\ \hline F & P
	\end{array}\right].
	\end{equation}
	It is shown in \cite[Lemma 1, pp. 7235]{tobenkin2010convex} that 
 \begin{equation}
\label{equiv:schur}
M \succ \gamma I, ~\gamma>0 ~ \iff ~ \mbox{Schur stability of} ~E^{-1}F.
 \end{equation}
Hence if there exist matrices $L$ and $R$ such that \eqref{eqn:A_param} and \eqref{eqn:M} hold, then $M \succeq \epsilon I$. Thus, $A(L, R)$ is Schur stable.
	
	($\Leftarrow$)
To prove necessity, invoking the equivalence in \eqref{equiv:schur}, it needs to be shown that a positive definite matrix $M\succ \gamma I$ can always be parameterized by $M = LL^\top + \epsilon I$ for some $L \in \rea^{2N\times 2N}$ and $\epsilon\in \rea_{>0}$. By the continuity of eigenvalues of a matrix with respect to its elements \cite[Ch. 7]{bhatia2013matrix}, one has that $M - \epsilon I$ is positive definite by choosing a sufficiently small positive $\epsilon \ll \gamma$. Therefore, the Cholesky factorization guarantees the existence of $L$ such that $M - \epsilon I = LL^\top$, as required.
 \hfill $\square$
\end{pf}

To represent the nonlinear function $\phi$ while maintaining sufficient flexibility, we propose the parameterization
\begin{equation}\label{eqn:phi_param}
\phi(x) = \col(x ,~ \varphi(x, \theta_{\tt NN}))
% \begin{bmatrix}
% x \\ \varphi(x, \theta_{\tt NN})
% \end{bmatrix}
\end{equation}
where the nonlinear part $\varphi$ can be any differentiable function approximator, parameterized by $\theta_{\tt NN}$. For brevity, the dependence on $\theta_{\tt NN}$ is dropped in the notation. Here, we choose $\varphi$ to be a feedforward neural network due to its scalability. The dimension $N$ is a hyperparameter chosen by the user. This specific structure guarantees the \emph{existence} of left inverses, though non-unique. We use a separate feedforward neural network 
$
 \phil =   \phil(\cdot, \theta_{\tt L})
$
with all unknown parameters collected in the vector $\theta_{\tt L}$ to approximate the inverse.%, which is shown to have better performance empirically.

\begin{rem}\rm 
Eq. \eqref{eqn:phi_param} is one of the feasible mappings described in Theorem \ref{thm:1}. In fact, the construction of \eqref{eqn:phi_param} can be further generalized as $\phi = \col(\psi, \varphi)$, where $\psi: \rea^n \to \rea^n$ is an invertible network, and $\varphi: \rea^n \to \rea^{N-m}$ is an arbitrary networks. Invertible neural networks have been extensively studied, with notable examples including  bi-Lipschitz networks \cite{WANetal}, invertible residual layers \cite{de2020block}, and monotone networks \cite{ahn2022invertible}. In this paper, we pick $\psi$ to be just the identity map for simplicity.
%
   % Note that there are many possible parameterizations to  observables that are compatible with the framework, with  In some specific parameterizations, the left inverse may be computed analytically and does not have to be modeled as a separate differential approximator. %For example, if $\phi(x) = [x^\top, \varphi(x)^\top]^\top $, then the left inverse is simply $x = C_0\phi(x)$ with $C_0 = [I, 0]$, though this may lose sufficient generality. 
\end{rem}

%==============
\subsection{Learning framework for Koopman embeddings}
\label{sec:43}
%==============

Under the parameterization of the proposed model class, we need to use the dataset $\cale_{\tt D}$ to fit the parameters 
$
\theta: = ( \theta_{\tt NN}, \theta_{\tt L}, L, R ).
$
% To this end, we need to solve optimization problems with a proper cost function. 
To this end, we consider minimizing the {simulation error} 
% \begin{equation} \label{eqn:sim_error}
$J_{\tt SE} := \frac{1}{T} \sum_{t=1}^{T} |\tilde{z}_{t} - {z}_t|^2$
% \end{equation}
in the lifted coordinate,
with $T = N_{\tt traj}$, $\tilde{z}_t = \phi(\tilde{x}_t)$, and $z_t = A(L,R)^{t-1} \phi(\tilde x_1)$. %Here, $\tilde{x}_t$ denotes the measured data corresponding to the true state $x_t$. 
To identify the left inverse $\phil$ concurrently, we minimize the composite cost function
\begin{equation} \label{eqn:min_J_se}
\hat\theta = \arg\min_{\theta \in \Theta} \; J_{\tt SE} + \alpha J_{\tt RE}
\end{equation}
with the parameter space $\Theta$, a weighting coefficient $\alpha>0$, and the reconstruction loss
$
J_{\tt RE} := \frac{1}{T} \sum_{t=1}^T |\tilde x_t - \phil(\phi(\tilde x_t, \theta_{\tt NN}), \theta_{\tt L})|^2.
$
The loss $J_{\tt RE}$ is used to learn a left-inverse $\phil$ for $\phi$, which can be thought of as a penalty term relaxing the constraint  $ x = \phil(\phi(x)) \ \forall x $. %, and the scalar $\alpha$ determines the weighting of the penalty. 
We can make full use of data from multiple trajectories by constructing the cost function as the sum of $(J_{\tt SE}^j + \alpha J_{\tt RE}^j)$, where, with a slight abuse of notations, the index $j$ denotes the data sets from different feasible trajectories. {It is beneficial to use multiple trajectories with different initial conditions to ensure the collected data is informative or ``sufficiently exciting'' \cite{LJU}. A sufficiently diverse set of initial conditions enhances system exploration and improves learning over a relatively large domain. However, establishing theoretical assumptions for nonlinear systems with guaranteed properties remains challenging, making it an interesting direction for both practical and theoretical exploration in future work.
}

\begin{rem}\rm\label{rem:error}
{Stability in learned dynamical models is often enforced via constrained optimization \cite{MamakoukasEtAl2020Memory,UMEMAN}. In \cite{MamakoukasEtAl2020Memory} stability is enforced iteratively by projecting the solution onto a feasible set; and  \cite{LIACOL} uses Gaussian processes in a probabilistic Koopman framework to quantify model uncertainty. In contrast, a standing advantage of the optimization \eqref{eqn:min_J_se} is its unconstrained formulation, while still guaranteeing stability. The set $\Theta$ is a Euclidean space.
%, and the mapping $\theta \mapsto \phi$ remains differentiable for commonly used parameterizations.
%, e.g. using the parameterization \eqref{eqn:phi_param} with $\varphi(\cdot, \theta_{\tt NN})$ as a neural network.
%
The resulting nonlinear model can be expressed as $x_{t+1} = \phil(A \phi(x_t))$. Due to learning errors, the identified mappings $\phi$ and $\phil$ may not satisfy $\phil \circ \phi = \mathbb{I}_d$. However, if they are locally Lipschitz, we have $|x_{t+1}| = |\phil(A \phi(x_t))|\le {\left| {\partial \phil \over \partial z} \right|} |z_{t+1}| $. Since $z_t \to 0$ as $t\to\infty$, it follows $|x_t|\to 0$. This ensures that even with imperfectly learned functions $\phi$ and $\phil$, the origin of the learned model is still attractive.}
\end{rem}

\begin{rem}
\rm
Minimizing simulation errors using finite data is widely popular in system identification \cite[Ch. 7]{LJU}. With noisy data, the estimates converge in expectation to their true values as the number of samples increases to infinity. %In practice, using sufficiently excited data from multiple trajectories improves accuracy. 
%As discussed above, despite parameter estimation errors, the imposed stability in our approach ensures the attractivity of the learned model. 
In Section \ref{sec:Koopman_invariance}, we provide an error analysis from a dynamical systems viewpoint for the case where the learned embedding $\phi$ is not Koopman invariant. Meanwhile, developing an error characterization with probabilistic bounds is an another interesting direction to explore.
\end{rem}

%It is worth emphasizing two important properties of Problem \eqref{eqn:min_J_se}. First, it is an \emph{unconstrained} optimization problem, but imposes guranteed stability on the identified lifted linear model. The parameter set $\Theta$ is the space of real numbers of the appropriate dimensionality. Second, there exists a differentiable mapping from the parameters $\theta$ to the objective for any choice of differentiable mapping $\phi(\cdot, \theta_{\tt NN})$, e.g. using the parameterization \eqref{eqn:phi_param} with $\varphi(\cdot, \theta_{\tt NN})$ as a neural network. Regarding the resulting nonlinear model, it can be expressed as $x_{t+1} = \phil(A \phi(x_t))$. Due to learning errors, the learnt mapings $\phi$ and $\phil$ may not satisfy $\phil \circ \phi = \mathbb{I}_d$. While it is not straightforward to verify stability of the identified nonlinear model in the $x$-coordinate, we can still ensure the attractivity of the equilibrium.

\begin{rem}\rm
The proposed approach enables the use of off-the-shelf first-order optimizer with autodiff toolboxs. This significantly simplifies the implementation, as autodiff software automatically computes gradients w.r.t. $\theta_{\tt NN}$ via the chain rule. Although the cost function in \eqref{eqn:min_J_se} is nonconvex, deep learning has proven effective in finding approximate local minima, instead of guaranteeing proximity to the global minimum. Besides, it allows flexibility in choosing differentiable cost functions beyond $J_{\tt SE} + \alpha J_{\tt RE}$. For instance, the simulation error in $x$ was tested but performed poorly, as a small error in $x$ does not ensure a well-structured embedding.
\end{rem}

\subsection{Koopman invariance \& error bound estimates}
\label{sec:Koopman_invariance}

{Our result in Theorem \ref{thm:1} establishes that, in theory, a class of stable nonlinear systems can be represented as finite-dimensional linear systems, the success of which is mainly built upon D1. In this subsection, we briefly discuss its connection to the Koopman invariance.
   
  The condition D1 explicitly imposes Koopman invariance. To see this, consider the set of mappings characterized by D1, i.e. 
   $
   \mathcal{G}:=\{\phi\in \mathcal{F} : {\phi\circ f}  - A \phi = 0, ~ \forall x \}.
   $
   It is straightforward to verify that 
   $$
   \phi\circ f \circ f- A \phi \circ f=0\; \implies\;\calk[\phi(x)] \in \mathcal{G},\; \forall \phi(x)\in \mathcal{G}.
   $$
   % since $\phi\circ f \circ f- A \phi \circ f=0$. 
   This directly satisfies Definition \ref{def:koopman_invariance}, confirming the Koopman invariance. For numerical implementation, the learned Koopman embedding $\phi$ may contain errors, i.e. 
   \begin{equation}
   \label{eq:epsilon}
   \phi\circ f(x) = A\phi(x) +\epsilon(x),
   \end{equation}
   with an error term $\epsilon$. Consequently, the resulting set of observables may no longer be Koopman invariant. The following provides an estimate of approximation errors.  

\begin{prop}
\rm 
For the nonlinear system \eqref{eq:syst} within any compact set $\cale \subset \rea^n$, if the learned Koopman embedding $\phi$ is not exact but satisfies \eqref{eq:epsilon}, then the norm of the prediction error in $z$ is bounded by ${C\epsilon_{\tt max} \over 1- \rho_{\tt A}}$, where $\rho_{\tt A} \in (0,1)$ is the spectral radius of $A$, $C:= \sup_{j \ge 0}{\|A^j\|_2 \over \rho^j_{\tt A}}< \infty$, and $\epsilon_{\tt max}:= \sup_{x\in \cale} |\epsilon(x)|$. 
\end{prop}
%%%
\begin{pf}
Defining $z=\phi(x)$ leads to the perturbed system $z_{t+1} = Az_t + \epsilon_t$, with $ \epsilon_t:=\epsilon(x(t))$. Meanwhile, the ``learned ideal'' lifted model becomes $\breve z_{t+1} = A\breve z_t$. In a compact set of interest $\cale$, we may find the upper bound $\|\epsilon_t\|_\infty\le \epsilon_{\tt max}$. 
   
The $z$- and $\breve z$-systems share the same initial condition, i.e.  $z(0) = \breve z(0) = \phi(x(0))$. The prediction error, arising from the breakdown of Koopman invariance, is given by $\breve z - z$, whose dynamics follows
\begin{equation}
  z_{t+1} -  \breve z_{t+1}   = A (z_t - \breve z_t) + \epsilon_t.
\end{equation}
Then, we have 
$$
\begin{aligned}
|z_t - \breve z_t|  &
\le \sum_{i=0}^{t-1} \|A^{t-1-i}\| \epsilon_{\tt max}
\le \epsilon_{\tt max}
\sum_{i=0}^{t-1} C\rho^{t-1-i}_{\tt A}
\\
& \le {C \|\epsilon_t\|_\infty \over 1- \rho_{\tt A}}
\end{aligned}
$$
for $C:=\sup_{j \ge 0} {\|A^j \|_2 \over \rho_{\tt A}^j}$ and $\rho_{\tt A}\in (0,1)$ due to the stability of $A$. From the Gelfand's formula $\lim_{j\to \infty}\|A^j\|_2^{1/j} = \rho_{\tt A}$, we conclude that $C$ is a bounded parameter.  \hfill $\square$ 
\end{pf}
   
   }

% \begin{rem}\rm
% The model class is agnostic to the particular cost function that is minimized. Unconstrained parameterization in the proposed framework has the benefit that $J_{\tt SE} + \alpha J_{\tt RE}$ may be replaced by other feasible differentiable objective functions. An alternative is the simulation error in the original $x$-coordinate. However, in practice, this was found to produce poor performance. The simulation error in $z$ can still be large while the simulation error in $x$ is small. As a result, the embedding may fit poorly without including the excess coordinates of $z$ in the minimization.
% \end{rem}
 
%==========

% \begin{rem}\rm
% In fact, the enforcement of stability in learned or identified dynamical models has been extensively studied, both within Koopman-based frameworks and beyond. Most approaches in the literature rely on constrained optimization, typically requiring specialized algorithms \cite{MamakoukasEtAl2020Memory,UMEMAN}. 
% %
% In \cite{MamakoukasEtAl2020Memory}, stability is enforced at every iteration step by projecting the solution onto the feasible set to handle the constraints. The recent work \cite{LIACOL} provides a probabilistic Koopman framework based on Gaussian processes, which quantifies model uncertainty in the context of learning dynamical models.
% %
% \end{rem}

%
%=======================
\section{Imitation Learning}
\label{sec:5}
%=======================
%

In this section, we address the motivating application P2 and extend the results in Section \ref{sec:4}. The main objective is to obtain a stabilizing controller that reproduces the demonstrated trajectories from a given plant. 
% In this paper, we propose to simultaneously learn a stabilizable model of the dynamics, which acts as a form of regularization encouraging closed-loop stability of the learned controller. 
We begin by proposing a class of stabilizable Koopman models.%, before turning to the problem of imitation learning.

\subsection{Stabilizable Koopman model class}

It is well-known that extending the Koopman operator to control systems is technically challenging and may yield \emph{bilinear} lifted systems \cite{GOSPAL,STRetal}. To obtain a \emph{bona fide} linear lifted model, we focus on a particular class of nonlinear systems, which are referred to as ``generalized feedback linearizable systems''.

\begin{defn}\rm 
    For the system \eqref{eq:syst}, if we can find mappings $\alpha:\rea^n \times \rea^m \to \rea^m$ and $\phi: \rea^n \to \rea^N$ ($N\ge n$), and the matrices $A \in \rea^{N\times N} $ and $ B\in \rea^{N\times m}$ satisfying     
 \begin{itemize}
     \item[C1:]  The algebraic equation
    \begin{equation}
    \label{eq:linear_cond}
   B^\bot [\phi\circ f_c (x,v) - A\phi (x) ] = 0, \quad \forall v \in \rea^m
    \end{equation}
with $f_c(x,v):= f(x) + g(x) \alpha(x,v)$;

\vspace{.1em}

    \item[C2:] The mapping $\phi$ is injective.
 \end{itemize}   
   Then we call the system \eqref{eq:syst} generalized feedback linearizable. In addition, if the pair $(A,B)$ is stabilizable, we refer to it as Koopman stabilizable.
\end{defn}

{After applying the pre-feedback $u=\alpha(x,v)$, the condition C1 yields $\phi(x_{t+1}) = \phi(f(x_t, v_t)) = A\phi(x_t) + Bv_t$. By defining the coordinate $z := \phi(x)$ and viewing $v$ as a new input, the control model becomes the LTI system
\begin{equation}
\label{syst:z}
z_{t+1} = A z_t + Bv_t,
\end{equation}
Since $\phi$ plays a similar role as in the Koopman embedding,  a similar role as in the Koopman embedding, we refer to the generalized feedback linearizable system as Koopman stabilizable if the pair $(A,B)$ is stabilizable.} 
% The injectivity condition C2 guarantees that the coordinate $z$ can be pulled back to the original $x$-coordinate.

\begin{rem}\rm \label{def:FL}
The above definition covers all feedback linearizable systems that involve a pre-feedback and a \emph{diffeomorphism} $\phi(\cdot)$, a concept central to constructive nonlinear control over the past three decades \cite{ISIbook,SASbook}. See \cite[Thm. 4.2.3]{ISIbook} for a necessary and sufficient condition of feedback linearizability and \cite{ARAetal} for a discrete-time version. In \cite{MONNOR}, this class of nonlinear systems is called ``immersed by feedback into a linear system'', and the authors provide a \emph{local} version of the necessary and sufficient condition via the differential geometric approach.\footnote{The necessary and sufficient condition in \cite{MONNOR} is proposed for \emph{affine} discrete-time systems. For ease of presentation, we also consider systems in the affine form \eqref{eq:syst}.} However, introducing the prefeedback term may prevent the establishment of connections between the nonlinear control system and the lifted linear system, preventing us from obtaining results similar to those in Theorem \ref{thm:1}.
\end{rem}

We now propose the stabilizable Koopman model class.

\begin{center}
    \fbox{ \parbox { 0.88\linewidth} 
{

\vspace{0cm}\small
{\bf M2.} Stabilizable Koopman Model ($A, B, \alpha, \phi, \phil$):
\begin{equation} \label{eqn:model2}
    \begin{aligned}
    z_0 & ~=~ \phi(x_0)\\
    z_{t+1} & ~=~ A z_t + Bv_t
    \\
    x_t & ~=~ \phil(z_t)
    \\
    u_t & ~=~ \alpha(x_t, v_t)
    \end{aligned}
\end{equation}
in which 1) $z\in \rea^N$ ($N \ge n$) is a lifted internal variable; 2) The pair $(A,B)$ is stabilizable; and 3) $\phi$ has a left inverse satisfying $\phil(\phi(x)) = x, \; \forall x$.
}
}
\end{center}

\subsubsection{Stabilization criterion}% for generalized feedback linearizable systems}

\begin{prop}\label{thm:stabilizability}\rm 
Assume the system \eqref{eq:syst} is Koopman stabilizable under the $C^1$-continuous pre-feedback $\alpha$ (w.r.t. $x$ and $v$) and an immersion $\phi: \rea^n \to \rea^N$, and $({\partial \phi \over \partial x})^\top {\partial \phi \over \partial x}$ is uniformly bounded. Then, any matrix $K$ that achieves Schur stability of $(A+BK)$ renders the closed-loop system $x_{t+1} = f_c(x_t,  K\phi(x_t))$ contracting.
\end{prop}

\vspace{-.2cm}
%%%%%%%%%%%%%%%%%%%%%%%%
\begin{pf} 
The system under $u = \alpha(x,v)$ becomes $x_{t+1} = f_c(x_t,v_t)$. Combining the above, the stabilizing feedback $v= K\phi(x)$ and \eqref{eq:linear_cond}, one gets
\begin{equation}
    \phi \circ f_c (x, K\phi) = (A+BK)\phi.
\end{equation}
Taking its partial derivative w.r.t. $x$, one gets
\begin{equation} \label{diff_relation}
\begin{aligned}
   \frac{\partial \phi}{\partial x}(x_{t+1})\left[ {\partial f_c \over \partial x}(x_t, \cdot)  - {\partial f_c \over \partial v}(x_t, \cdot)K\frac{\partial \phi}{\partial x}(x_t) \right] \\ = (A+ BK)\frac{\partial \phi}{\partial x}(x_t),
\end{aligned}
\end{equation}
where we have used the $C^1$-continuity of $\alpha(\cdot,\cdot)$.

The stabilizing controller for \eqref{syst:z} implies the existence of a matrix $P \succ 0$ such that
\begin{equation} \label{stabilisability}
    P - (A+BK)^\top P (A+BK) \succ I \succeq \rho I,
\end{equation}
with $\rho\in(0,1]$. It yields
\begin{equation} \label{stab2}
    \left(\frac{\partial \phi}{\partial x}\right)^\top [P - (A+BK)^\top P (A+BK)] \frac{\partial \phi}{\partial x} \succ \rho\left(\frac{\partial \phi}{\partial x}\right)^\top  \frac{\partial \phi}{\partial x}.
\end{equation}
We consider a candidate metric
$
    M(x) := {\partial \phi \over \partial x}(x)^\top P {\partial \phi \over \partial x}(x),
$
and substitute \eqref{diff_relation} into \eqref{stab2}, obtaining $\beta  := \mbox{$ {\rho \over \lambda_{\tt max} (P)}$}$ and
\begin{equation*}
\begin{aligned}
    M(x_t) - {\partial f_x \over \partial x}(x_t)^\top M(x_{t+1}) {\partial f_x \over \partial x}(x_t) & \succ 
    \rho\frac{\partial \phi}{\partial x}
    (x_t)^\top  \frac{\partial \phi}{\partial x}(x_t)
    \\
    & \succeq \beta M(x_t)
\end{aligned}
\end{equation*}
in which, for convenience, we have defined a new function $f_x(x) := f_c(x, K\phi(x))$. By choosing $\rho \in (0, 1]$ sufficiently small, we have $0< \beta < 1$. Therefore, the closed-loop dynamics $x_{t+1} = f_c(x_t, K\phi(x_t))$ is contracting.
\hfill $\square$
\end{pf}

The matching equation \eqref{eq:linear_cond} is closely connected to the condition in control contraction metrics (CCM), originally proposed in \cite{manchester2017control}, and a continuous-time version of the connection between CCM and Koopman stabilizability is revealed in \cite[Sec. VI-A]{YiManchester2021equivalence}. In the above proposition, the additional smoothness assumption of the prefeedback term $\alpha$ facilitates contraction analysis. While this may appear somewhat conservative, it is a common and practical scenario to consider smooth feedback for a given smooth model.

In general, finding the mappings $\alpha$ and $\phi$ analytically is non-trivial. In Section \ref{sec:52}, we explore how to simultaneously learn these mappings purely from data. Since we only have access to ``local'' data from the region in which the plant is operated, the learning approach is inherently effective only within this local domain. Consequently, this implies that the system is required to satisfy the generalized feedback linearization condition only locally within the region of interest -- rather than globally -- for practical implementation.

\subsubsection{Parameterizing Koopman stabilizable systems}

In \textbf{M2}, we need to identify four components: a stabilizable pair $(A,B)$, a pre-feedback $\alpha$, the mapping $\phi$, and its left inverse $\phil$. To facilitate the learning framework, in this section, we study how to parameterize them.

Earlier works on learning controllers have used linear matrix inequality (LMI) constraints to impose stabilizability \cite{havens2021imitation,yin2021imitation}. However, the computation of constrained optimization becomes extremely expensive when jointly estimating the system dynamics. In the following, an \emph{unconstrained} parameterization of the triple $(A, B, K)$ is proposed with a guaranteed stabilizable pair $(A,B)$.

\begin{prop} \label{prop:stabilizable}\rm 
Consider a pair $(A, B) \in \rea^{N\times N} \times \rea^{N\times m}$ with $\mbox{rank}\{B\} = m$ and $A$ parameterized as %$A(\theta_{\tt SL})$
	\begin{equation} \label{eqn:A_param2}
    \begin{aligned}
	A(\theta_{\tt SL})&
 = 
 \begin{bmatrix}
     B^\bot \\ B^\top
 \end{bmatrix}^{-1}
 \begin{bmatrix}
	    2 B^\bot (M_{11} + M_{22} + R - R^\top)^{-1}M_{21} \\ S
	\end{bmatrix},
 \\ 
\theta_{\tt SL} &:= (L,R,S, B)
    \end{aligned}
	\end{equation}
	where $M_{ij}$ ($i,j =1,2$) are blocks in 
	\begin{equation} \label{eqn:M2}
	M := \begin{bmatrix} M_{11} & M_{12} \\ M_{21} & M_{22} \end{bmatrix} = LL^\top + \epsilon I,
	\end{equation}
	with $\epsilon>0$, $L \in \rea^{2N \times 2N}$, and $R \in \rea^{N\times N}$. Then, the pair $(A,B)$ is stabilizable. Conversely, for any stabilizable pair $(A, B)$, we can always find $\theta_{\tt SL}$ and $\epsilon$ to parameterize it in the form of \eqref{eqn:A_param2}.
\end{prop}

\begin{pf}
The stabilizability of the pair $(A,B)$ is equivalent to the existence of a matrix $K \in \rea^{m \times n}$ such that $A_{\tt CL}: = A+BK$ is Schur stable. Invoking Proposition \ref{prop:stable_param}, it is necessary and sufficient to have matrices $L, R$ and $\epsilon$ to parameterize the closed-loop system matrix $A_{\tt CL}$ as 
    \begin{equation}
    \label{para_cl}
        A_\texttt{CL} = 2 (M_{11} + M_{22} + R - R^\top)^{-1}M_{21}.
    \end{equation}
From
 $ \mbox{rank}  \{ \col( B^\bot, B^\top)\} = n,
 % \begin{bmatrix}
 %        B^\bot \\ B^\top
 %    \end{bmatrix} = n,
    $
we multiply to \eqref{para_cl}, thus
\begin{equation*}
    \begin{bmatrix}
        B^\bot \\ B^\top
    \end{bmatrix}
    A
    =
    \begin{bmatrix}
      2 B^\bot (M_{11} + M_{22} + R - R^\top)^{-1}M_{21}
      \\
      S
    \end{bmatrix}    
\end{equation*}
    with 
    $
     S: =    - B^\top BK +   2 B^\top (M_{11} + M_{22} + R - R^\top)^{-1}M_{21}.
    $
Considering $\mbox{rank} \{B^\top B\} = m$ and the freedom of $K$, hence $S$ is a free variable to parameterize $A$. Since all the above implications are necessary and sufficient, we complete the proof. \hfill $\square$
\end{pf}

\begin{rem}\rm 
In the parameterization, the sub-block $M_{22}$ qualifies as a Lyapunov matrix $P$ due to $M_{22} - (A + BK)^\top M_{22}(A+BK) \succ 0$, in which case $K = \frac{1}{2}B^\top P A_\texttt{CL}$. On the other hand, there are infinite numbers of feasible selections for $K$ to guarantee Schur stability.
\end{rem}

In the control case, the mapping $\phi$ shares the same properties as the one in Section \ref{sec:4}, and thus we adopt the same parameterizations of $\phi$ and its left inverse $\phil$ as done in Section \ref{sec:412}. The nonlinear function $\alpha$ can be parameterized using another neural network. Notably, we found that an invertible mapping from $v$ to $u$ is beneficial in specific training, and we present empirical results from simulations in Section \ref{sec:6}.

\subsection{Imitation learning framework}
\label{sec:52}

We now consider the imitation learning (IL) of stabilizable Koopman models to learn a controller that reproduces trajectories of the system \eqref{eq:syst} demonstrated by an expert policy, given the data set $\mathcal{E}_{\tt D}:=\{\tilde x_t, \tilde u_t\}_{t=1}^{\nt}$. One well-studied and widely-used paradigm for IL frames it as a supervised learning problem and directly fits a mapping from state to control input. This is referred to as behavioral cloning \cite{bain1995framework}, which aims to minimize the cost function $J_{\tt BC} =    \min_{\theta \in \Theta} \sum_{i=1}^{N_{\tt traj}}  |\tilde u^i_t - k_\theta(\tilde x^i_t)|^2 + r(\theta)$, with $r(\cdot)$ a regularization function. %Here, we study to learn stabilizing controllers using the model class \textbf{M2}. 

Different from some works with known dynamics \cite{havens2021imitation,tu2022sample,yin2021imitation}, in this paper stability is used to regularize IL with unknown dynamics. Our approach is to jointly learn a stabilizable model and a stabilizing controller. 
We make use of the model set {\bf M2} and the data set $\cale_{\tt D}$ to estimate the parameters
$
    \theta:= (L,R, S, B, \theta_{\tt NN}, \theta_{\tt L}).
$
Similar to the autonomous case, we solve the \emph{unconstrained} optimization containing the simulation error and a stability regularization penalty term:

\begin{equation} \label{eqn:min2}
\hat\theta = \arg\min_{\theta \in \Theta}  \Big( c_1 J_{\tt SE}' + c_2 J_{\tt SL} +  c_3 \alpha J_{\tt RE}'\Big)
\end{equation}
where we define $J_{\tt SE}' = \sum_{t=1}^{T-1} |\tilde z_{t+1} - A \tilde z_{t} - B\tilde v_t |^2$, $J_{\tt SL} = \sum_{t=1}^{T-1} |\tilde z_{t+1} - A_\texttt{CL}\tilde z_{t}|^2 + \left|\tilde v_t - \half  B^\top PA_\texttt{CL}\tilde z_t \right|^2$, and $J_{\tt RE}' =  \sum_{t=1}^T |\tilde x_t - \phil(\phi(\tilde x_t))|^2 $,
% \begin{equation}
% \begin{aligned}
% J_{\tt SE}' & = \sum_{t=1}^{T-1} |\tilde z_{t+1} - A \tilde z_{t} - B\tilde v_t |^2
% \\
% J_{\tt SL} & = \sum_{t=1}^{T-1} |\tilde z_{t+1} - A_\texttt{CL}\tilde z_{t}|^2 + \left|\tilde v_t - \half  B^\top PA_\texttt{CL}\tilde z_t \right|^2
% \\
% J_{\tt RE}' & =  \sum_{t=1}^T |\tilde x_t - \phil(\phi(\tilde x_t))|^2 
% \end{aligned}
% \end{equation}
with $T= N_{\tt traj}$, $A_{\tt CL}= A+ BK$, weighting coefficients $c_i>0$ ($i=1,2,3$), and $\tilde z_t = \phi(\tilde x_t)$ and $\tilde v_t = \alpha^{-1}(\tilde x_t, \tilde u_t)$, invoking the bijectivity of $\alpha$.

% that the selected function $\alpha(\cdot, \cdot)$ is bijective. 

The term $J_{\tt SE}'$ is the ``open-loop'' simulation error in the lifted coordinate, i.e. treating $\tilde v$ as an exogenous input, over stabiliziable pairs $(A, B)$; $J_{\tt SL}$ can be viewed as the simulation error in the lifted coordinate in closed-loop, over Schur-stable matrices $A_{\tt CL}$, along with a term similar to behavioural cloning for $\tilde v$; finally, the last term $J_{\tt RE}'$ is to ensure the left invertibility of $\phi$ and learn its left inverse. This optimization is also solvable via autodiff software.

\begin{rem}
\rm
The smoothness assumption of $\phi$ and the prefeedback simplify contraction analysis while also facilitating the use of autodiff toolboxes for learning. However, smoothness is somewhat conservative for discrete-time systems. In the continuous-time case, it excludes systems that do not satisfy Brockett’s necessary condition \cite{BRO}. Nevertheless, despite this requirement, the proposed approach remains applicable to a broad class of practical systems. 
On the other hand, an approximate $\alpha$ may introduce a mismatch between the nonlinear system and its lifted one. If the mismatch is small, it can be treated as a perturbation and analyzed using standard techniques \cite{khalil2002nonlinear}, in which case the closed loop may remain exponentially stable. However, larger mismatches lead to steady-state errors or instability. Studying its practical impact is an interesting direction for future research.
\end{rem}

%
%=======================
\section{Simulation Results}
\label{sec:6}
%=======================
%

\subsection{Learning stable Koopman embedding}

\begin{figure}[!htp]
    \centering
    \begin{subfigure}{0.32\textwidth}
        \centering
        \includegraphics[width=\textwidth]{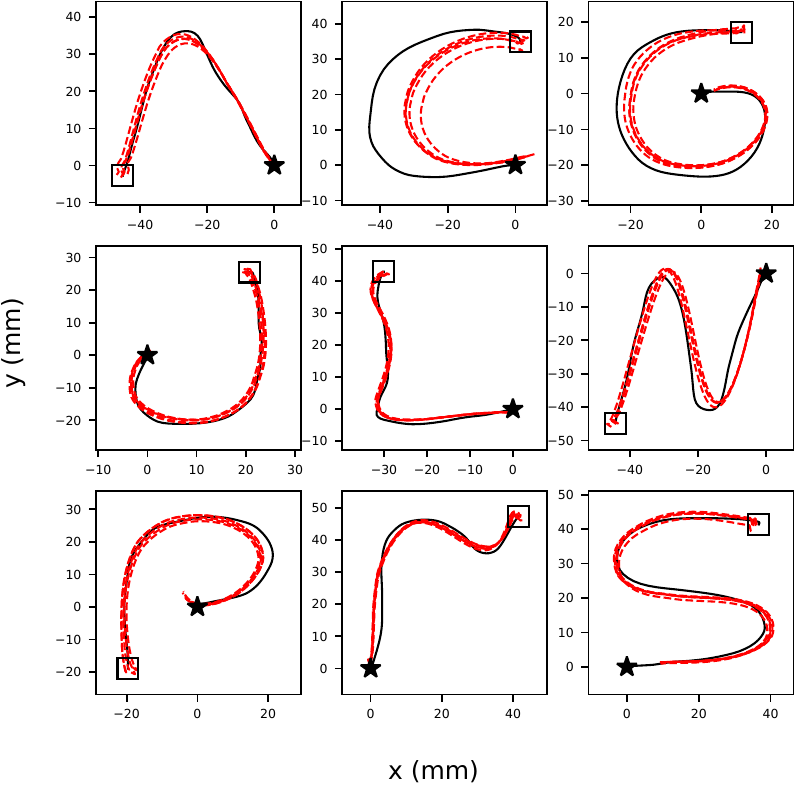}
        
        \vspace{-.2cm}
        \caption{SKEL (ours)}
        \label{fig:skel_sim}
    \end{subfigure}
    %\hfill
    % \hspace{0.07\textwidth}%
    \begin{subfigure}{0.32\textwidth}
        \centering
        \includegraphics[width=\textwidth]{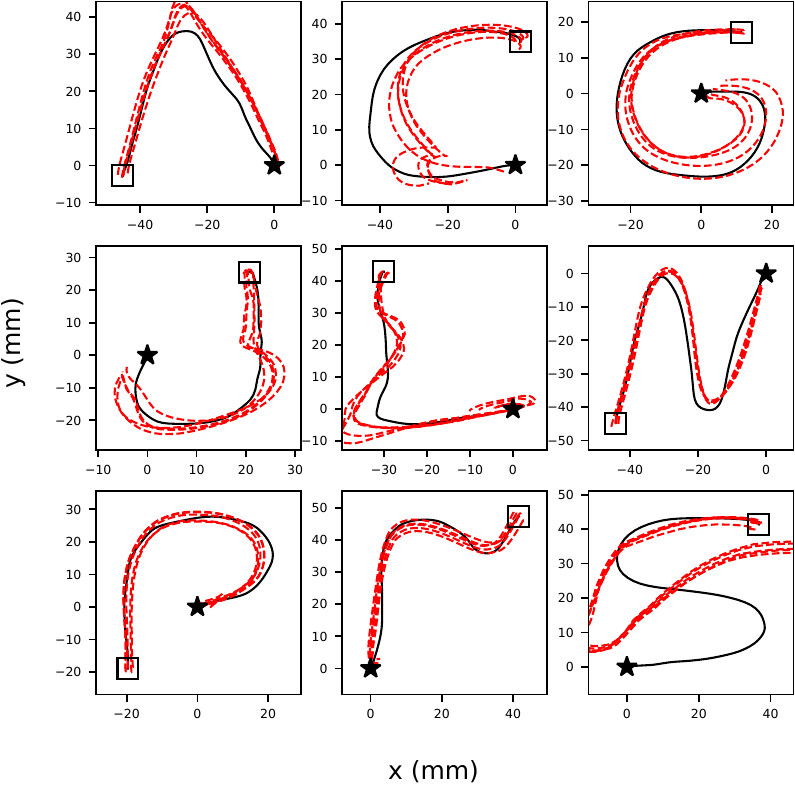}

        \vspace{-.2cm}
        \caption{LKIS \cite{TakeishiEtAl2017Learning}}
        \label{fig:lkis_sim}
    \end{subfigure}
\caption{Simulations of SKEL and LKIS models on test data. (Red dotted lines: trajectories from the models; Solid black line: true trajectories with the endpoint denoted by $\star$; Initial conditions are sampled from the square region.)
% Trajectories from the models are shown as red dotted lines. True trajectory is shown as a solid black line, with the endpoint denoted by $\star$. 
}
\label{fig:lasa_simulations}
\end{figure}

We validated the approach in Section \ref{sec:4} on the LASA handwriting dataset \cite{KhansariBillard2011Learning,SCHetal}, which consists of human-drawn trajectories of various letters and shapes.
% \footnote{https://cs.stanford.edu/people/khansari/download.html}.
In the handwriting example, the motions can be modeled as second-order stable systems, satisfying the contraction assumption in compact sets. Moreover, as the sampled data discretize a continuous-time system, the vector field $f$ is invertible.
%The dataset has been widely used as a benchmark for learning stable systems \cite{SCHetal}. %Stability is an important constraint for the system characterized by this dataset as unconstrained models can have spurious attractors, leading to poor generalization to unseen initial conditions.
A discrete-time model was trained for each shape in the dataset to follow a desired path from any initial condition. To prepare the data, splines were fitted to the trajectories and the datapoints were re-sampled at a uniform time interval. The state was chosen as $\tilde x_t = [y_t^\top, \dot y_{t}^\top]^\top$, where $y_t , \dot y_t \in \rea^2$ are the position and velocity vectors of the end-effector.
%formulating the minimal realization to this system. 
All data was scaled to the range $[-1, 1]$ before training. For each shape, leave-one-out cross validation was performed. Test trajectories are plotted in Fig. \ref{fig:skel_sim}.
% as solid black lines for a subset of the shapes in the dataset. 
The proposed learning framework was implemented in PyTorch
%\footnote{https://github.com/pytorch/pytorch} 
and the {ADAM} optimizer \cite{KINBA} was used to solve the optimization \eqref{eqn:min_J_se}. Hyperparameter values were chosen to be $\alpha = 10^3$ and $\epsilon = 10^{-8}$. All instances of $\varphi$ were selected as fully-connected feedforward neural networks using rectified linear units (ReLU) as the activation function with its parameter $b$, 2 hidden layers with 50 nodes each, and an output dimensionality of 20. The neural network parameters $\theta_{\tt NN}$ and $\theta_{\tt L}$ are initialized using the default scheme in PyTorch, while $L$, $R$, and $b$ are initialized randomly from a uniform distribution. For simplicity, our proposed framework is denoted as SKEL. %(Stable Koopman Embedding Learning). 

\begin{figure*}[!htp]
    \centering
    \begin{subfigure}[t]{0.3\textwidth}
        \centering
      \includegraphics[width=1\textwidth]{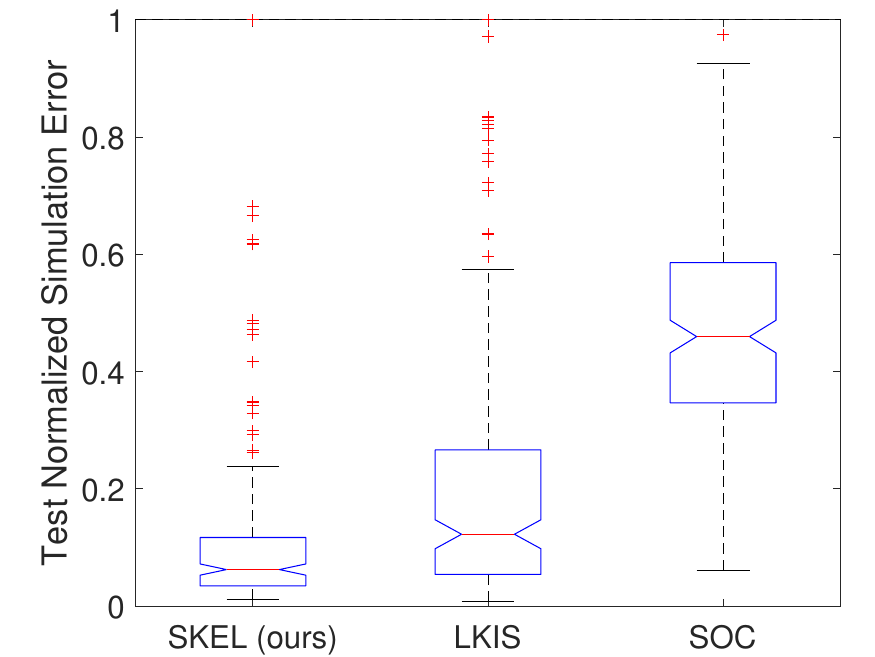}
         \caption{}
        \label{fig:skel_boxplot}
    \end{subfigure}
    \hfill
    \begin{subfigure}[t]{0.3\textwidth}
        \centering
    \includegraphics[width=1\textwidth]{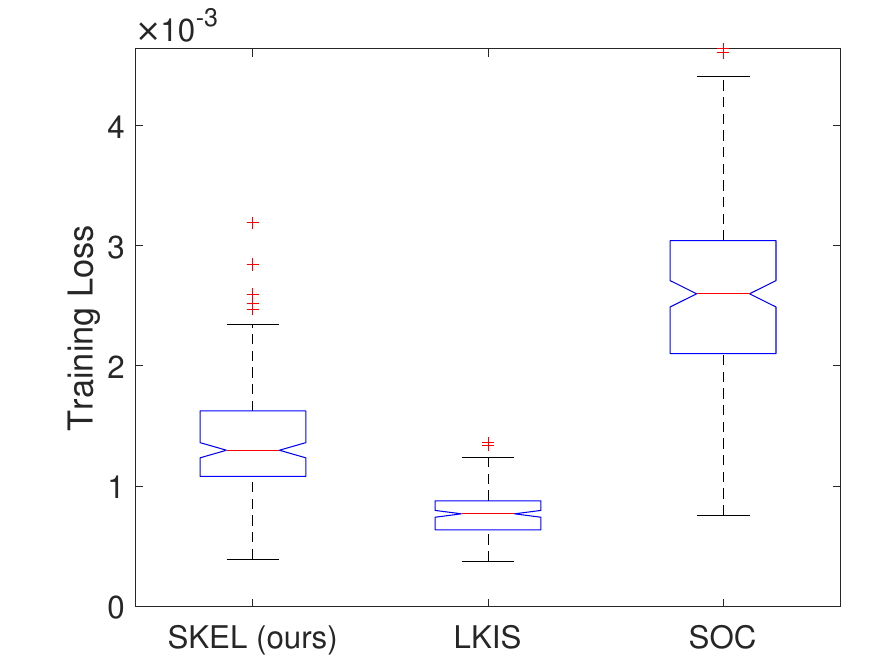}
    \caption{}
    \label{fig:skel_training}
    \end{subfigure}
    \hfill
    \begin{subfigure}[t]{0.3\textwidth}
        \centering
    \includegraphics[width=1\textwidth]{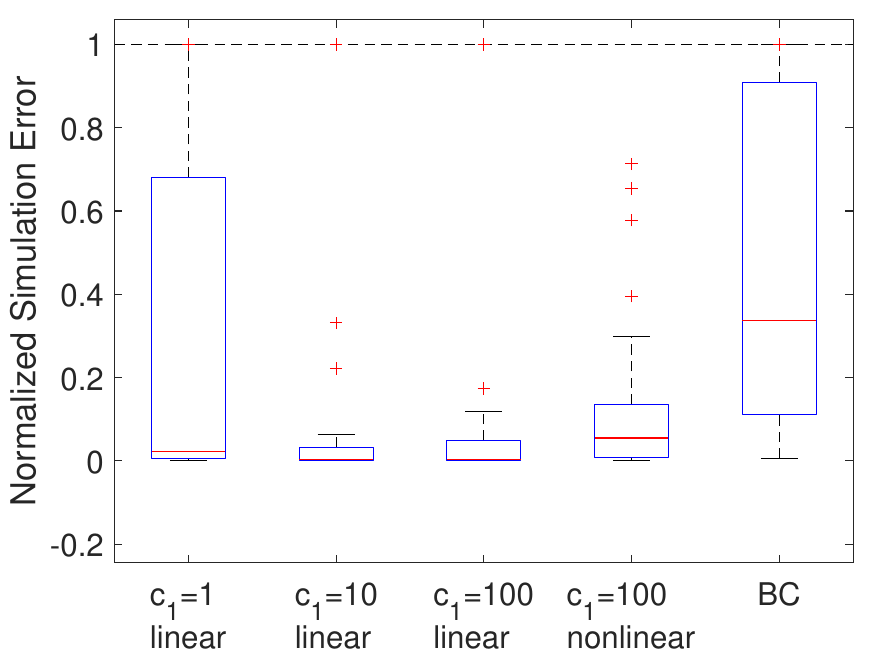}
    \caption{}
    \label{fig:lasa_imitation_nse}
    \end{subfigure}
\caption{Boxplots: (a) Comparison of SKEL with other Koopman learning methods. Outliers were clipped for better visibility of boxes. Number of outliers with NSE $>1$ from left to right: 1 (SKEL), 15 (LKIS), 0 (SOC); (b) Training loss for each method; (c) Normalized simulation error of learned controllers on the test set. From left to right: linear parameterization of $\alpha$ --- $c_1=1$, $c_1=10$ and $c_1=100$, nonlinear parameterization of $\alpha$ --- $c_1=100$, behavioural cloning (BC). Number of clipped outliers from left to right: 4, 2, 1, 0, 5.}
\end{figure*}

We compared with a constrained stable parameterization (SOC) in \cite{MamakoukasEtAl2020Memory} and an unconstrained parameterization (LKIS) in \cite{TakeishiEtAl2017Learning} which does not have stability guarantees. We kept most aspects the same when fitting different models, using the normalized simulation error (NSE) 
$
{\tt NSE} = ({\sum_{t=1}^T  | \hat x_t - \tilde x_t|^2})/(
{\sum_{t=1}^T |\tilde x_t|^2}),
$
where $\{\hat x\}^T_{t=1} $ is the simulated trajectory using the learned model. A boxplot of the normalized simulation error for the three methods is shown in Fig. \ref{fig:skel_boxplot}. Our method achieves the lowest median NSE on the test set with 95\% confidence. From Fig. \ref{fig:skel_training}, LKIS attains the lowest training error, but does not generalize to the test set and SKEL. This can be interpreted as a symptom of overfitting and shows that the stability guarantees of SKEL have a regularizing effect on the model. With regards to SOC, it was observed with relatively high training and test errors. A qualitative evaluation was performed to determine the robustness of the models to small perturbations in the initial condition of the test trajectory. Only SKEL and LKIS were compared as it was clear from Fig. \ref{fig:skel_boxplot} that SOC underperformed in this setting. The results are plotted in Figs. \ref{fig:skel_sim}-\ref{fig:lkis_sim}. It can be seen that SKEL produces trajectories that converge to each other,
%due to their contractivity, 
whereas the LKIS models sometimes behave unpredictably.
% , indicating instability of the learned model.

% \begin{figure}
%     \centering
%     \includegraphics[width=0.28\textwidth]{Figures/boxplot_clip.pdf}
%     \caption{Comparison of SKEL with other Koopman learning methods. Outliers were clipped for better visibility of boxes. Number of outliers with NSE $>1$ from left to right: 1 (SKEL), 15 (LKIS), 0 (SOC).}
%     \label{fig:skel_boxplot}
% \end{figure}

% \begin{figure}
%     \centering
%     \includegraphics[width=0.28\textwidth]{training_error_boxplot.pdf}
%     \caption{Training loss for each method}
%     \label{fig:skel_training}
% \end{figure}

\subsection{Imitation learning} \label{sec:62}
% {\em Planar Robotic Manipulator.} 
The approach in Section \ref{sec:5} was validated on the same handwriting trajectories.
% to be imitated. 
The data were generated by a simulated 2 degree-of-freedom (DoF) robot, whose dynamics at the end-effector can be simplified as fully-actuated if we are only concerned with the working space rather than the configuration space. It has an Euler-Lagrange form and is \emph{feedback linearizable}, thus satisfying the key assumptions. Since the LASA dataset only contains state trajectories, an inverse dynamics model was used to generate torques as control inputs for imitation learning; see \cite[Sec. 5]{fan2023learning} for further details.
% . Further details on the model used are given in .

\begin{figure}[!htp]
    \centering
    \begin{subfigure}{0.32\textwidth}
        \centering
        \includegraphics[width=\textwidth, trim=0 3 0 0, clip]{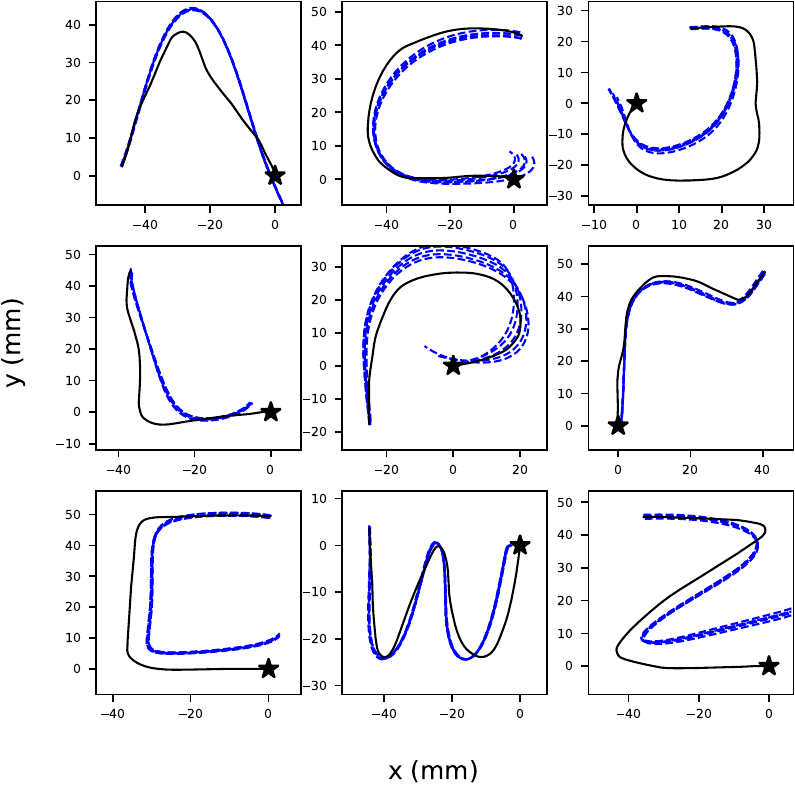}

        \vspace{-.2cm}
        \caption{Our method (linear $\alpha$, $c_1 = 100$)}
        \label{fig:skel_imitation}
    \end{subfigure}
    %\hfill
    \hspace{0.07\textwidth}%
    \begin{subfigure}{0.32\textwidth}
        \centering
        \includegraphics[width=\textwidth, trim=0 3 0 0, clip]{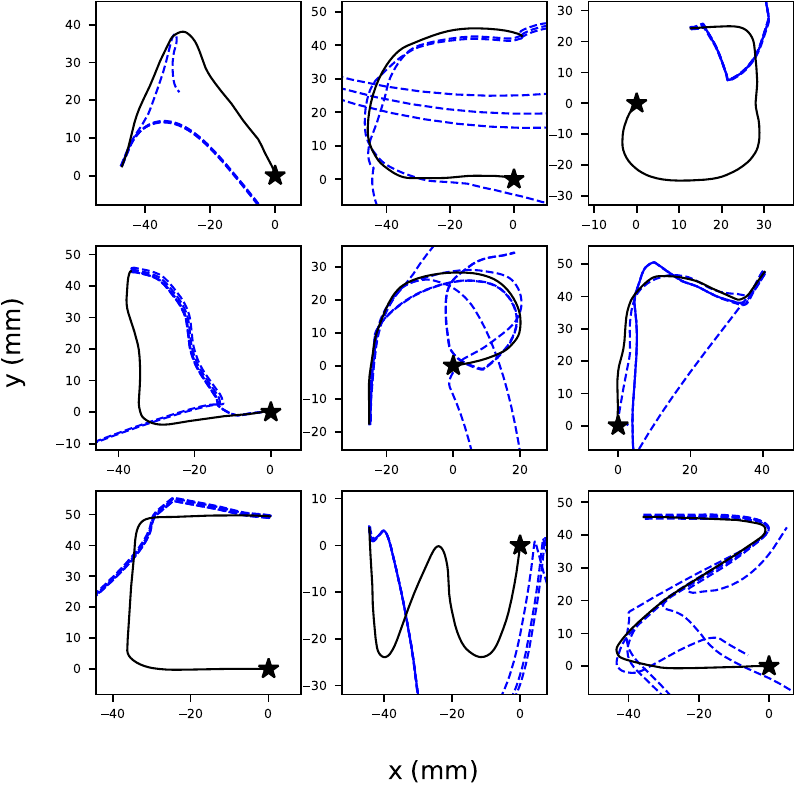}

        \vspace{-.2cm}
        \caption{Behavioural cloning}
        \label{fig:bc_imitation}
    \end{subfigure}
\caption{Simulations of the closed loop. (Blue dotted lines: trajectories produced by the controllers; Solid black line: true trajectories with the endpoint denoted by the star.)
% Trajectories produced by the controllers are shown as blue dotted lines. The true trajectory is shown as a solid black line, 
}
\label{fig:imitation_simulations}
\end{figure}

Comparisons were made of the performance of the learned controller for various values of $c_1$, and also made against the standard behavioral cloning (BC) method \cite{fu2020d4rl}. 
% which is commonly used as a baseline for evaluating imitation learning algorithms \cite{fu2020d4rl}.
BC was implemented as fitting a neural network mapping states to control inputs by minimizing a mean-squared error loss on the controller output. The neural networks were chosen to have 2 hidden layers with 20 nodes and tanh activations. For a quantitative comparison, NSE was used with 
$
    \sum_{t=1}^{T-1} \left. {|\tilde x_t - \hat x_t}|^2 / \sum_{t=1}^{T-1}|\tilde x_t|^2 \right. ,
$
 where $\hat x_{t+1} = f(\hat x_{t}) + g(\hat x_t)k(\hat x_t)$ and $\hat x_1 = \tilde x_1$. Regarding the prefeedback function $\alpha$, we consider two bijective choices: 1) $u=v$ having been used in learning-based control \cite{han2020deep,kaiser2021data,KordaMezic2018Linear}; 2) an affine coupling layer \cite{DinhEtAl2017Density}:
    $u = v\odot \exp(s(x)) + h(x),$
where $\odot$ denotes the Hadamard product, and $s, h$ can be arbitrary function approximators. The second choice includes the nonlinear function \(h\), which adds flexibility to model the nonlinearity in \(\alpha\), while having an analytical inverse
$
    v = (u - h(x)) \odot \exp(-s(x)).
$
Bijectivity is particularly useful for enhancing training performance.%\footnote{There are various methods for learning invertible maps, e.g. invertible residual networks \cite{de2020block} and BiLipNet \cite{WANetal}.}
% \begin{figure}
%     \centering
%     \includegraphics[width=0.28\textwidth]{Figures/lasa_imitation_boxplot.pdf}
%     \caption{Normalized simulation error of learned controllers on the test set. From left to right: linear parameterization of $\alpha$ --- $c_1=1$, $c_1=10$ and $c_1=100$, nonlinear parameterization of $\alpha$ --- $c_1=100$, behavioural cloning (BC). Number of clipped outliers from left to right: 4, 2, 1, 0, 5.}
%     \label{fig:lasa_imitation_nse}
% \end{figure}
Fig. \ref{fig:lasa_imitation_nse} shows the NSE for the learned controllers. Note that increasing $c_1$ reduces the NSE up to a point, beyond which performance deteriorates. Besides, increasing the model complexity by using the nonlinear parameterization for $\alpha$ does not reduce the NSE, possibly because the small training dataset was insufficient for training larger models. Meanwhile, the BC approach has a substantially larger NSE than the best-performing controller from our proposed method, which shows that the proposed stability-based regularization does indeed improve control performance.
% of the controller over the baseline. 
A comparison of the trajectories produced by the learned controllers is shown in Fig. \ref{fig:imitation_simulations}. The controller produced by our method induces a closed-loop in which nearby trajectories remain close to each other and converge to a single equilibrium, as expected for contracting systems, whereas the trajectories of the BC controller results in divergent trajectories even with small perturbations to the initial condition.
% , which is unacceptable when controlling physical systems. 
Our approach provides an obvious improvement over BC for the same requirements on the data and without a significant increase in computational cost, and has a regularizing effect on learning stabilizing controllers.

% The results suggest that the proposed approach does have a regularizing effect on learning stabilizing controllers.
% and outperforms the BC method in terms of imitation error. 

\begin{figure}[!htp]
    \centering
    \includegraphics[width=0.3\textwidth]{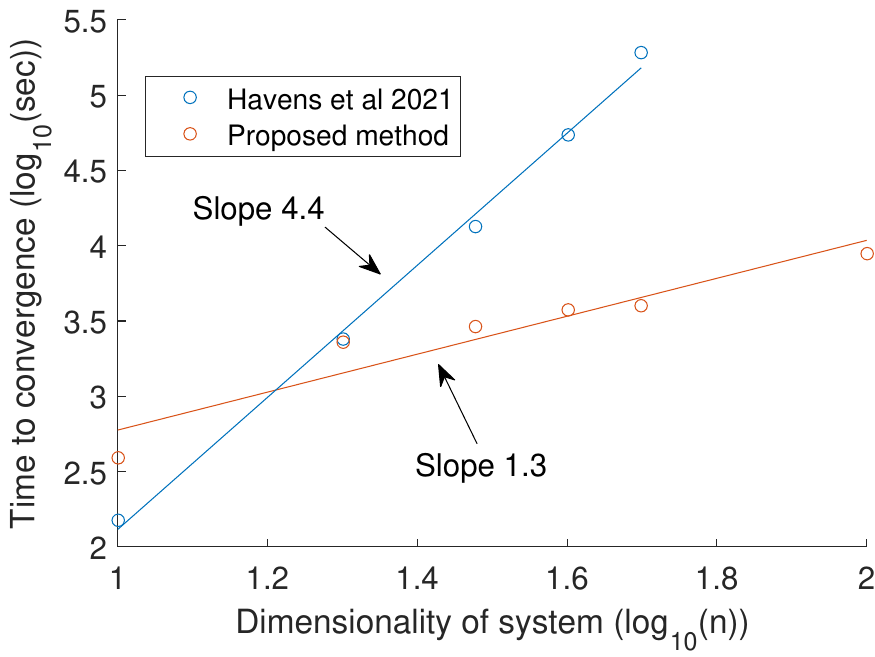}
    \caption{Scatter plot of total time to convergence of the proposed method vs. the PGD algorithm of \cite{havens2021imitation} in log-log scale, plus lines of best fit.}
    \label{fig:linear_time_to_converge}
\end{figure}

To evaluate the benefits of direct parameterization, we investigated scalability using a linear example \cite{dean2020sample} which models an unstable graph Laplacian system. We artificially generated trajectory data for learning, with more details in \cite[Sec. 5]{fan2023learning}. A comparison was made against a prior stability-constrained imitation learning method \cite{havens2021imitation} that requires exact knowledge of $(A,B)$. Their method was applied by first estimating $(A, B)$ via least squares.\footnote{Note that the method of estimating the open-loop dynamics in \cite{havens2021imitation} is not applicable or extensible to nonlinear systems.} The scalability of both algorithms was evaluated by measuring computation time to convergence of the optimization problems. For the proposed method, this corresponds to the time taken to compute the gradient and update the parameters. The projected gradient descent (PGD) algorithm proposed by \cite{havens2021imitation} requires solving a semidefinite program at each iteration. Fig. \ref{fig:linear_time_to_converge} shows the total convergence time. The slopes of the lines of best fit reveal how the computation times scale with the dimensionality of the system. It can be seen that the proposed method is substantially more scalable thanks to the unconstrained model. 

%============================
%============================

\section{Conclusion}
We introduced new classes of Koopman models with stability and stabilizability guarantees, which are built upon our novel theoretical connections between the contraction and Koopman stability criteria in the paper. The stable Koopman model has been applied to nonlinear system identification, while the stabilizable Koopman model class has shown efficacy in solving imitation learning. In both cases, we proposed parameterization methods to obtain \emph{unconstrained} optimization problems to significantly reduce computation burden. Testing on the renowned LASA handwriting dataset demonstrated that our approaches outperform pervious methods lacking stability guarantees.

\begin{ack}                               % Place acknowledgements
This work was supported in part by the Australian Research Council (ARC), the Natural Sciences and Engineering Research Council of Canada (NSERC), and the Programme PIED of Polytechnique Montr\'eal. The authors would like to thank the four anonymous reviewers for their thorough review and valuable feedback. 
\end{ack}

\bibliographystyle{abbrv}    
\bibliography{thesis} 

\begin{thebibliography}{10}

\bibitem{ahn2022invertible}
B.~Ahn, C.~Kim, Y.~Hong, and H.~J. Kim.
\newblock Invertible monotone operators for normalizing flows.
\newblock {\em Adv. Neural Inf. Process. Syst.}, 35:16836--16848, 2022.

\bibitem{ARAetal}
E.~Aranda-Bricaire, {\"U}.~Kotta, and C.~H. Moog.
\newblock Linearization of discrete-time systems.
\newblock {\em SIAM J. Control Optim.}, 34(6):1999--2023, 1996.

\bibitem{bain1995framework}
M.~Bain and C.~Sammut.
\newblock A framework for behavioural cloning.
\newblock In {\em Mach. Intell.}, pages 103--129, 1995.

\bibitem{bhatia2013matrix}
R.~Bhatia.
\newblock {\em Matrix Analysis}.
\newblock Springer, 1996.

\bibitem{BRIetal}
L.~Brivadis, V.~Andrieu, and U.~Serres.
\newblock Luenberger observers for discrete-time nonlinear systems.
\newblock In {\em IEEE Conf. Decis. Control}, pages 3435--3440. IEEE, 2019.

\bibitem{BRO}
R.~Brockett.
\newblock Asymptotic stability and feedback stabilization.
\newblock {\em Differ. Geom. Control Theory}, 27(1):181--191, 1983.

\bibitem{DAVetal}
A.~Davydov, A.~V. Proskurnikov, and F.~Bullo.
\newblock Non-{E}uclidean contractivity of recurrent neural networks.
\newblock In {\em Am. Control Conf.}, pages 1527--1534, 2022.

\bibitem{de2020block}
N.~De~Cao, W.~Aziz, and I.~Titov.
\newblock Block neural autoregressive flow.
\newblock In {\em Uncertain. Artif. Intell.}, pages 1263--1273, 2020.

\bibitem{dean2020sample}
S.~Dean, H.~Mania, N.~Matni, B.~Recht, and S.~Tu.
\newblock On the sample complexity of the linear quadratic regulator.
\newblock {\em Found. Comput. Math.}, 20(4):633--679, 2020.

\bibitem{DEV}
R.~L. Devaney.
\newblock {\em A First Course in Chaotic Dynamical Systems: Theory and Experiment}.
\newblock CRC Press, 1992.

\bibitem{DinhEtAl2017Density}
L.~Dinh, J.~Sohl{-}Dickstein, and S.~Bengio.
\newblock Density estimation using real {NVP}.
\newblock In {\em Int. Conf. Learn. Represent.} OpenReview.net, 2017.

\bibitem{fan2023learning}
F.~Fan.
\newblock {\em Learning Stable {Koopman} Models for Identification and Control of Dynamical Systems}.
\newblock PhD thesis, 2023.
\newblock The University of Sydney.

\bibitem{Fan2022Learning}
F.~Fan, B.~Yi, D.~Rye, G.~Shi, and I.~R. Manchester.
\newblock Learning stable {K}oopman embeddings.
\newblock In {\em Proc. Am. Control Conf.}, pages 2742--2747, 2022.

\bibitem{FICBIL}
B.~Fichera and A.~Billard.
\newblock Learning dynamical systems encoding non-linearity within space curvature.
\newblock {\em arXiv preprint arXiv:2403.11948}, 2024.

\bibitem{fu2020d4rl}
J.~Fu, A.~Kumar, O.~Nachum, G.~Tucker, and S.~Levine.
\newblock D4rl: {D}atasets for deep data-driven reinforcement learning.
\newblock {\em arXiv preprint arXiv:2004.07219}, 2020.

\bibitem{GillisEtAl2020note}
N.~Gillis, M.~Karow, and P.~Sharma.
\newblock A note on approximating the nearest stable discrete-time descriptor systems with fixed rank.
\newblock {\em Appl. Numer. Math.}, 148:131--139, 2020.

\bibitem{GOSPAL}
D.~Goswami and D.~A. Paley.
\newblock Bilinearization, reachability, and optimal control of control-affine nonlinear systems: A koopman spectral approach.
\newblock {\em IEEE Trans. Autom. Control}, 67(6):2715--2728, 2021.

\bibitem{han2020deep}
Y.~Han, W.~Hao, and U.~Vaidya.
\newblock Deep learning of {K}oopman representation for control.
\newblock In {\em IEEE Conf. Decis. Control}, pages 1890--1895. IEEE, 2020.

\bibitem{havens2021imitation}
A.~Havens and B.~Hu.
\newblock On imitation learning of linear control policies: Enforcing stability and robustness constraints via {LMI} conditions.
\newblock In {\em Am. Control Conf.}, pages 882--887, 2021.

\bibitem{ISIbook}
A.~Isidori.
\newblock {\em Nonlinear Control Systems}.
\newblock Springer, 3 edition, 1995.

\bibitem{kaiser2021data}
E.~Kaiser, J.~N. Kutz, and S.~L. Brunton.
\newblock Data-driven discovery of {K}oopman eigenfunctions for control.
\newblock {\em Mach. Learn. Sci. Technol.}, 2(3):035023, 2021.

\bibitem{khalil2002nonlinear}
H.~K. Khalil.
\newblock {\em Nonlinear Systems}.
\newblock Prentice Hall, 2002.

\bibitem{KhansariBillard2011Learning}
S.~M. Khansari-Zadeh and A.~Billard.
\newblock Learning stable nonlinear dynamical systems with {G}aussian mixture models.
\newblock {\em IEEE Trans. Robot.}, 27(5):943--957, 2011.

\bibitem{KINBA}
D.~P. Kingma and J.~Ba.
\newblock {ADAM}: A method for stochastic optimization.
\newblock In {\em Int. Conf. Mach. Learn.}, 2015.

\bibitem{Koopman1931Hamiltonian}
B.~O. Koopman.
\newblock Hamiltonian systems and transformation in {H}ilbert space.
\newblock {\em Proc. Natl. Acad. Sci.}, 17(5):315--318, 1931.

\bibitem{KordaMezic2018Linear}
M.~Korda and I.~Mezi\'{c}.
\newblock Linear predictors for nonlinear dynamical systems: {K}oopman operator meets model predictive control.
\newblock {\em Automatica}, 93:149--160, 2018.

\bibitem{KrizhevskyEtAl2012Imagenet}
A.~Krizhevsky, I.~Sutskever, and G.~E. Hinton.
\newblock Imagenet classification with deep convolutional neural networks.
\newblock In {\em Adv. Neural Inf. Process. Syst.}, pages 1097--1105. Curran Associates, Inc., 2012.

\bibitem{lacy2003subspace}
S.~L. Lacy and D.~S. Bernstein.
\newblock Subspace identification with guaranteed stability using constrained optimization.
\newblock {\em IEEE Trans. Autom. Control}, 48(7):1259--1263, 2003.

\bibitem{LANMEZ}
Y.~Lan and I.~Mezi{\'c}.
\newblock Linearization in the large of nonlinear systems and {Koopman} operator spectrum.
\newblock {\em Phys. D.}, 242(1):42--53, 2013.

\bibitem{LIACOL}
Y.~Lian and C.~Jones.
\newblock On {Gaussian} process based {Koopman} operators.
\newblock {\em IFAC-PapersOnLine}, 53(2):449--455, 2020.

\bibitem{LJU}
L.~Ljung.
\newblock {\em System Identification: Theory for the User}.
\newblock Prentice Hall PTR, 2 edition, 1999.

\bibitem{lohmiller2000control}
W.~Lohmiller and J.-J. Slotine.
\newblock Control system design for mechanical systems using contraction theory.
\newblock {\em IEEE Trans. Autom. Control}, 45(5):984--989, 2000.

\bibitem{LohmillerSlotine1998Contraction}
W.~Lohmiller and J.-J.~E. Slotine.
\newblock On contraction analysis for non-linear systems.
\newblock {\em Automatica}, 34(6):683--696, 1998.

\bibitem{MamakoukasEtAl2020Learning}
G.~Mamakoukas, I.~Abraham, and T.~D. Murphey.
\newblock Learning stable models for prediction and control.
\newblock {\em IEEE Trans. Robot.}, 39(3):2255--2275, 2023.

\bibitem{MamakoukasEtAl2020Memory}
G.~Mamakoukas, O.~Xherija, and T.~Murphey.
\newblock Memory-efficient learning of stable linear dynamical systems for prediction and control.
\newblock In {\em Adv. Neural Inf. Process. Syst.}, pages 13527--13538. Curran Associates, Inc., 2020.

\bibitem{manchester2017control}
I.~R. Manchester and J.-J.~E. Slotine.
\newblock Control contraction metrics: Convex and intrinsic criteria for nonlinear feedback design.
\newblock {\em IEEE Trans. Autom. Control}, 62(6):3046--3053, 2017.

\bibitem{Manek2019Learning}
G.~Manek and J.~Z. Kolter.
\newblock Learning stable deep dynamics models.
\newblock In {\em Adv. Neural Inf. Process. Syst.}, volume~32. Curran Associates, Inc., 2019.

\bibitem{mania2022active}
H.~Mania, M.~I. Jordan, and B.~Recht.
\newblock Active learning for nonlinear system identification with guarantees.
\newblock {\em J. Mach. Learn Res.}, 23(1):1433--1462, 2022.

\bibitem{mauroy2016global}
A.~Mauroy and I.~Mezi{\'c}.
\newblock Global stability analysis using the eigenfunctions of the {K}oopman operator.
\newblock {\em IEEE Trans. Autom. Control}, 61(11):3356--3369, 2016.

\bibitem{MITetal}
M.~Mitjans, L.~Wu, and R.~Tron.
\newblock Learning deep {K}oopman operators with convex stability constraints.
\newblock {\em ArXiv Preprint}, 2024.
\newblock arXiv:2404.15978.

\bibitem{MONNOR}
S.~Monaco and D.~Normand-Cyrot.
\newblock The immersion under feedback of a multidimensional discrete-time non-linear system into a linear system.
\newblock {\em Int. J. Control}, 38(1):245--261, 1983.

\bibitem{PanDuraisamy2020Physics}
S.~Pan and K.~Duraisamy.
\newblock Physics-informed probabilistic learning of linear embeddings of nonlinear dynamics with guaranteed stability.
\newblock {\em SIAM J. Appl. Dyn. Syst.}, 19(1):480--509, 2020.

\bibitem{revay2023recurrent}
M.~Revay, R.~Wang, and I.~R. Manchester.
\newblock Recurrent equilibrium networks: Flexible dynamic models with guaranteed stability and robustness.
\newblock {\em IEEE Trans. Autom. Control}, 69(5):2855--2870, 2023.

\bibitem{SANPRA}
R.~G. Sanfelice and L.~Praly.
\newblock Convergence of nonlinear observers on $\mathbb{R}^n$ with a riemannian metric (part i).
\newblock {\em IEEE Trans. Autom. Control}, 57(7):1709--1722, 2011.

\bibitem{SASbook}
S.~Sastry.
\newblock {\em Nonlinear Systems: Analysis, Stability, and Control}.
\newblock Springer, 1999.

\bibitem{SCHetal}
M.~Schonger and et~al.
\newblock Learning barrier-certified polynomial dynamical systems for obstacle avoidance with robots.
\newblock In {\em IEEE Int. Conf. Robot. Autom.}, pages 17201--17207, 2024.

\bibitem{singh2021learning}
S.~Singh, S.~M. Richards, V.~Sindhwani, J.-J.~E. Slotine, and M.~Pavone.
\newblock Learning stabilizable nonlinear dynamics with contraction-based regularization.
\newblock {\em Int. J. Robot. Res.}, 40(10-11):1123--1150, 2021.

\bibitem{STRetal}
R.~Str{\"a}sser, M.~Schaller, K.~Worthmann, J.~Berberich, and F.~Allg{\"o}wer.
\newblock {SafEDMD}: A certified learning architecture tailored to data-driven control of nonlinear dynamical systems.
\newblock {\em ArXiv Preprint}, 2024.
\newblock (arXiv:2402.03145).

\bibitem{sze2017efficient}
V.~Sze, Y.-H. Chen, T.-J. Yang, and J.~S. Emer.
\newblock Efficient processing of deep neural networks: A tutorial and survey.
\newblock {\em Proc. IEEE}, 105(12):2295--2329, 2017.

\bibitem{TakeishiEtAl2017Learning}
N.~Takeishi, Y.~Kawahara, and T.~Yairi.
\newblock Learning {K}oopman invariant subspaces for dynamic mode decomposition.
\newblock In {\em Adv. Neural Inf. Process. Syst.}, pages 1130--1140, 2017.

\bibitem{tobenkin2010convex}
M.~M. Tobenkin, I.~R. Manchester, J.~Wang, A.~Megretski, and R.~Tedrake.
\newblock Convex optimization in identification of stable non-linear state space models.
\newblock In {\em IEEE Conf. Decis. Control}, pages 7232--7237. IEEE, 2010.

\bibitem{TRABER}
G.~Q.~B. Tran and P.~Bernard.
\newblock Arbitrarily fast robust {KKL} observer for nonlinear time-varying discrete systems.
\newblock {\em IEEE Trans. Autom. Control}, (3):1520--1535, 2024.

\bibitem{tu2022sample}
S.~Tu, A.~Robey, T.~Zhang, and N.~Matni.
\newblock On the sample complexity of stability constrained imitation learning.
\newblock In {\em Learn. Dyn. Control Conf.}, pages 180--191, 2022.

\bibitem{UMEMAN}
J.~Umenberger and I.~R. Manchester.
\newblock Specialized interior-point algorithm for stable nonlinear system identification.
\newblock {\em IEEE Trans. Autom. Control}, 64(6):2442--2456, 2019.

\bibitem{WANetal}
R.~Wang, K.~Dvijotham, and I.~R. Manchester.
\newblock Monotone, bi-{L}ipschitz, and {Polyak-Lojasiewicz} networks.
\newblock In {\em Int. Conf. Mach. Learn.} JMLR.org, 2024.
\newblock Art. no. 2062.

\bibitem{YiManchester2021equivalence}
B.~Yi and I.~R. Manchester.
\newblock On the equivalence of contraction and {K}oopman approaches for nonlinear stability and control.
\newblock {\em IEEE Trans. Autom. Control}, 69(7):4336--4351, 2024.

\bibitem{yi2021reduced}
B.~Yi, R.~Wang, and I.~Manchester.
\newblock Reduced-order nonlinear observers via contraction analysis and convex optimization.
\newblock {\em IEEE Trans. Autom. Control}, 67(8):4045--4060, 2021.

\bibitem{yin2021imitation}
H.~Yin, P.~Seiler, M.~Jin, and M.~Arcak.
\newblock Imitation learning with stability and safety guarantees.
\newblock {\em IEEE Control Syst. Lett.}, 6:409--414, 2021.

\end{thebibliography}

\end{document}